\documentclass[12pt,letterpaper]{article}

\usepackage[includeheadfoot,
            marginratio={1:1,2:3}, 
            width=412pt,
            height=688pt,]{geometry}
            \usepackage{tikz}
\usepackage[compat=1.1.0]{tikz-feynman}
\usepackage[utf8]{inputenc}
\usepackage{amsmath}
\usepackage{amsfonts}
\usepackage{amssymb}
\usepackage{graphicx}
\usepackage{booktabs}
\usepackage{empheq}
\usepackage{paralist}
\usepackage{cite}
\usepackage[hidelinks]{hyperref}
\usepackage{xcolor}
\usepackage{tensor}
\usepackage{adjustbox}
\usepackage{subcaption}
\usepackage{braket}
\usepackage{yfonts}
\usetikzlibrary{decorations.pathreplacing,calc}

\newcommand{\nc}{\newcommand}
\nc{\lb}{\llbracket}
\nc{\rb}{\rrbracket}
\nc{\gl}{\llbracket}
\nc{\gr}{\rrbracket}
\nc{\bbR}{\mathbb{R}}
\nc{\bbC}{\mathbb{C}}
\nc{\bbZ}{\mathbb{Z}}
\nc{\cO}{\mathcal{O}}
\nc{\cS}{\mathcal{S}}
\nc{\cM}{\mathcal{M}}
\nc{\cT}{\mathcal{T}}
\nc{\cX}{\mathcal{X}}
\nc{\cQ}{\mathcal{Q}}
\nc{\cD}{\mathcal{D}}
\nc{\cC}{\mathcal{C}}
\nc{\cF}{\mathcal{F}}
\newcommand\beq{\begin{equation}}
\newcommand\eeq{\end{equation}}
\nc{\del}{\partial}
\nc{\tri}{\hspace{-3.5pt}\vartriangle\hspace{-3.5pt}}
\nc{\blacktri}{\blacktriangle}
\nc{\eq}[1]{\begin{equation}
                     \begin{split} #1 \end{split}
                     \end{equation}}
\nc{\ul}{\underline}
\nc{\ov}{\overline}
\nc{\fa}{\hat}
\nc{\fb}{\MakeUppercase}
\nc{\fc}{\tilde }
\nc{\Lie}{{\cal L}} 
\nc{\lambdabar}{{\mkern0.75mu\mathchar '26\mkern -9.75mu\lambda}}

\newmuskip\pFqmuskip

\newcommand*\pFq[7][8]{
  \begingroup 
  \pFqmuskip=#1mu\relax
  \mathchardef\normalcomma=\mathcode`,
  \mathcode`\,=\string"8000
  \begingroup\lccode`\~=`\,
  \lowercase{\endgroup\let~}\pFqcomma
  {}_{#2}{#3}_{#4}{\left[\left.\genfrac..{0pt}{}{#5}{#6}\right|#7\right]}
  \endgroup
}
\newcommand{\pFqcomma}{{\normalcomma}\mskip\pFqmuskip}

\DeclareMathOperator\arctanh{arctanh}

\allowdisplaybreaks[2]

\tikzset{/tikzfeynman/warn luatex=false}

\begin{document}

\vspace*{1.5cm}
\begin{center}
{\huge
The Tameness of Quantum Field Theory \\[0.1in]
\large
Part I -- Amplitudes
}
\vspace{.6cm}
\end{center}

\vspace{0.35cm}
\begin{center}
 Michael R.~Douglas,$^{1,2}$
 Thomas W.~Grimm,$^{3}$ and
 Lorenz Schlechter$^{3}$
\end{center}

\vspace{.5cm}
\begin{center} 
\vspace{0.25cm} 
\emph{$^{1}$Center of Mathematical Science and Applications, Harvard University, USA}\\
\emph{$^{2}$
Department of Physics, YITP and SCGP, Stony Brook University, USA}\\
\emph{$^{3}$
Institute for Theoretical Physics, Utrecht University,
The Netherlands } \\
   
\vspace{0.2cm}

\vspace{0.3cm}
\end{center}

\vspace{0.5cm}


\begin{abstract}

We propose a generalized finiteness principle for physical theories, 
in terms of the concept of tameness in mathematical logic.
A tame function or space can only have a finite amount of structure, in a precise sense which we explain.
Tameness generalizes the notion of an analytic function to include certain non-analytic
limits, and we show that this includes many limits which are known to arise in physics. 

For renormalizable quantum field theories, we give a general proof that amplitudes 
at each order in the loop expansion are tame functions of the external momenta and the couplings.   
We then consider a variety of exact non-perturbative results and show that they are tame but
only given constraints on the UV definition of the theory. 
This provides further evidence for the recent conjecture of the second author
that all effective theories that can be coupled to quantum gravity are tame. 
We also discuss whether renormalization group flow is tame, and comment on the applicability 
of our results to effective theories. 

\end{abstract}

\clearpage

\tableofcontents


\newpage

\parskip=.2cm
\section{Introduction}

Consider a physical theory and its observables -- trajectories $x(t)$ in classical mechanics,
expectation values in quantum mechanics, or correlation functions in a field theory.
The observables are functions, for example, of time, of positions or momenta, or any parameter of the theory, and
a natural mathematical question is to characterize the class of functions to which they belong.
Although very abstract, this information might be far easier to get than finding the specific functions. Furthermore, it can
provide general insights about the theory that lead to new ways to determine or constrain these functions. A general characterization of this type can also open the door for the classification of physical theories and provide a systematic 
way to show that two theories are actually different. 

This very general idea can take many forms.  Probably the most studied case is the question of whether a classical mechanical system is integrable.  Another very familiar idea is to characterize the singularities
of the observables, or their rate of growth at infinity.  The axioms of axiomatic quantum field theory
include conditions on correlation functions of this type (regularity, linear growth) which can be shown 
to imply more familiar physical conditions such as causality.  More recently the study of 
supersymmetric effective field theory makes heavy use of constraints on singularities of particle
masses and central charges as functions of moduli, obtained by combining supersymmetry constraints
with consistency of weakly coupled limits.

We often think of these properties as geometric, meaning that the properties in question can
be formulated so that they remain invariant under reparameterizations of the independent variables
(space and time coordinates, moduli space coordinates, {\it etc.}).  Breaking this invariance down,
it generally has two parts.  One which is familiar from general relativity is that many geometric
quantities are not scalars, and transform covariantly under reparameterizations.  Any geometric
property must behave consistently under these transformations, {\it i.e.}~be covariant.  But the other
requirement is that the functions in question, when composed with the functions
expressing the reparameterization, must remain in the same class.  For example, if we are interested
in holomorphic functions on field space, the interesting reparameterizations will also be defined by
holomorphic functions.  This is not a question of covariance {\it per se} but rather of consistently
working with functions of a particular class.

In mathematics there are many more function classes with nice behavior under composition and
other algebraic operations (addition, multiplication, {\it etc.}).  A much used example is
the class of $C^k$ real functions, meaning functions whose derivatives (in each variable) up
to total order $k$ exist and are continuous.  By the chain rule, these are closed under
composition.  One can define a ``manifold with $C^k$ structure'' by requiring that the charts
are $C^k$ functions, and in this sense this is also a geometric property.
A $C^\infty$ or ``smooth'' function has continuous derivatives of all orders, while an analytic
function is one which given any point, can be expressed as a power series which converges in a
neighborhood of that point.  As will be familiar to all mathematicians and most physicists,
there are many real smooth functions which are not analytic, indeed knowing the values of a
smooth function $f(x)$ for all $x\le 0$ (with $x\in\bbR$) says nothing about its values for $x>0$.
Conversely if $f(x)$ were analytic, the Taylor series at $x=0$ would have a finite radius
of convergence, determining the function up to some $a>0$.  By repeating this process, we could
break up any compact interval into a finite set of subintervals, in each of which we have an
explicit convergent series expansion for $f(x)$.

This finiteness property implies in particular that an equation such as $f(x)=0$ can have only
finitely many solutions on a compact interval.  This is not to say that a real analytic function
can only have a finite number of zeroes, consider for example $\sin x$.  However, an infinite
number of zeroes (as in this case) can only arise on a noncompact domain.  Similarly,
the function $\sin (1/x)$, sometimes called the topologist's sine function, is not analytic on the closed interval $[0,1]$ despite remaining bounded there.

Those functions, which on a compact domain can only have a finite number of zeroes or a finite amount of other structure, one can loosely refer to as ``tame'' functions. 
Here the term is used to contrast with ``wild'' examples from set theory and topology
such as the Cantor set, the Sierpinski gasket, the Weierstrass function, the topologist's sine function and so forth.  These wilder sets and functions will not concern us here. From a physicist's perspective, it might not come as a surprise that they are not needed to describe physics. However, the point is that introducing a \textit{precise notion of tameness}, as we will do in this work, and excluding all wild behavior gives a remarkably non-trivial constraint on a physical theory.

At this point the reader may be asking: yes, this is so, but are you not simply saying that the
functions in question are real analytic, a familiar concept that did not require this long review.
No.  While real analytic functions on a compact domain are tame, the surprising fact is
that there are larger classes of tame functions, which can be defined on domains that are non-compact, and simple physical reasons to consider these classes. 

Consider for example the partition function $Z(g)$ of
scalar $\phi^4$ quantum field theory (QFT) 
as a function of the coupling constant $g$.  This is a
function on $g\in[0,\infty)$.  It might even have a $g\rightarrow\infty$ limit, but to get 
started let us choose some $0<g_{\rm max}\ll 1$ and ask whether $Z(g)$ is tame on $[0,g_{\rm max}]$.  We know
a lot about the small $g$ behavior from perturbation theory and semiclassical methods. 
It turns out that $Z(g)$ is not real analytic at 0.  In contrast, physical intuition 
says that a QFT becomes simple at weak coupling, so it is a very reasonable 
supposition that $Z(g)$ is tame on this compact interval. 
Evidently this could only be true with some broader definition of
tameness.
Another example, from supersymmetric QFT and string theory, concerns the behavior of the effective action on moduli space.  As we mentioned earlier, these have highly constrained singularities, and a simple example is the $S \log S$ term in the Veneziano-Yankielowicz superpotential of $d=4$, $N=1$ super Yang-Mills theory.  This example can be realized by many string theory constructions, and there are many generalizations of it to gauge theories with matter, exotic theories and so on.
Is there a definition of ``tame function'' which includes everything which can come out of supersymmetric QFT and string theory, or even every possible theory which can be coupled to quantum gravity?

The class of tame functions that will concern us in this work are defined by using the 
tame geometry built from o-minimal structures \cite{VdDbook}. 
The starting point is very basic; 
one begins by first defining a novel type of topology, a `tame topology'. While o-minimal structures were introduced in mathematical logic \cite{VdDbook,Marker1996ModelTA,MacW} to define axiomatic logical systems 
that can be studied without settling hard logic questions, e.g.~about decidability, the associated tame topology realizes Grothendieck's dream for having a `topology for geometers' \cite{Esquisse} without the pathologies that arise in general set theory. We will define tame topologies in \S \ref{definability}, 
but the main point is that they are rich enough to allow for many 
non-trivial functions, such as the real exponential $e^x$ on the entire real line, the logarithm $\log x$ on the positive real axis, or analytic functions restricted to an interval.
Thus they have a chance of including functions such as the $\phi^4$ partition function,
the superpotentials of effective $D=4$ gauge theory, and many more. One
o-minimal structure containing many non-trivial functions is known as $\bbR_{\rm an,exp}$,
and it will be central in many of our applications. However, note that there are many 
more such structures and all share the strong finiteness properties imposed by o-minimality.

Let us close this introduction by recalling that tameness first entered physics through the study of the finiteness of flux vacua in string theory \cite{Bakker:2021uqw}. In fact, the flux superpotentials arising in string compactifications, such as the GVW superpotential \cite{Gukov:1999ya}, are tame functions of the moduli as discussed in \cite{Grimm:2021vpn}. The observation that tameness appears to be a common 
feature of all string theory effective actions was then promoted to a general `Tameness conjecture' in \cite{Grimm:2021vpn}. As we will review, this conjecture asserts that all effective actions that admit a UV completion with quantum gravity are tame in a 
well-defined sense. Our study of the tameness of quantum field theories at the perturbative and non-perturbative level will further elucidate this conjecture. 

In part II of this work \cite{Douglas:2023fcg}, we will propose new tameness conjectures for spaces of
effective field theories and conformal field theories.  

\subsection{Summary of results} \label{summary}

\subsubsection*{Feynman amplitudes are tame}

A main result of the present work is to show that Feynman amplitudes (for a very broad class of QFT's) are tame functions of the particle masses, coupling constants, and momenta (or equivalently the Lorentz invariant quantities formed from these).  Thanks to recent mathematical work on tameness and Hodge theory,
our argument can be very concise: we will show that amplitudes are period integrals, and call on a theorem
of Bakker and Mullane  \cite{bakker2022definable} following \cite{bakker2020tame,bakker2020definability} which states that such integrals are definable in the o-minimal structure $\bbR_{\rm an,exp}$. This brings us to the following result. 

\noindent \textbf{Theorem.} For any renormalizable quantum field theory with finitely many particles and interactions all amplitudes with finitely many loops and external legs 
are definable in the o-minimal structure 
$\bbR_{\rm an,exp}$ as functions of the masses, external momenta, and coupling constants.

We note that the renormalizability condition can be relaxed if one requires that the Lagrangian and the required counterterms at $n$-loop level are polynomial in the fields 
and derivatives. 

\subsubsection*{Tameness at non-perturbative level and in effective theories}

Studying simple quantum field theories with known partition functions we are able to establish 
several tameness results. We discuss the 0d sine-Gordon model, quantum mechanics, 2d linear sigma models, 2d Yang-Mills theory, and 3d non-critical M-theory. 
In these examples we show various tameness results: the partition functions, when known explicitly, are
$\bbR_{\rm an,exp}$-definable when viewed as functions of the coupling constants. These results follow by exploiting the relation of the partition functions to period integrals. 
Remarkably, we find that even in the simplest 0d $\phi^4$-theories  exponential periods arise that are not covered by the definability results of \cite{bakker2020tame,bakker2020definability,bakker2022definable}.

To investigate the tameness of the renormalization group we highlight recent mathematical 
insights on the interplay of first order differential equations and o-minimal structures. 
In particular, we stress that tameness is in conflict with the existence of RG limit cycles. 
In contrast, tameness is naturally preserved 
when integrating out heavy fields, both classically and with finite loop corrections. This 
suggests that tameness might be generally preserved under lowering the cutoff scale. 

It is not expected that every QFT is tame at the non-perturbative level and we elucidate a 
number of ways how tameness can be violated. 
In particular, we discuss the relation of tameness to the absence of global symmetries and 
highlight how tameness is easily violated if the UV theory admits a non-tame behavior. 
Combining these arguments, we collect further support for the conjecture 
\cite{Grimm:2021vpn} that tameness might 
be related to the consistency of an effective theory with quantum gravity.

\section{Tame geometry and o-minimal structures}
\label{definability}

In this section we give a brief introduction to tame geometry and o-minimal structures. 
The starting point is to define a tame topology of $\bbR^n$, which has strong 
finiteness properties, and serves as a foundation to define tame manifolds, tame functions, and 
other tame geometric objects \cite{VdDbook}.

\subsubsection*{Defining tame sets and functions}

The fundamental object underlying a tame geometry is an o-minimal structure $\cS$ that collects subsets 
of each $\bbR^n$, $n>0$. These sets are called $\cS$-definable, or \textit{definable} for short. To 
introduce this concept, let us first define a 
\textit{structure}. Denote by $\mathcal{S}_n$ a collection of subsets of $\bbR^n$. 
The structure $\cS=\{\cS_n\}_{n=1,2,...}$ should be closed under simple operations that can be performed among sets and should be sufficiently rich. The axioms for a structure are:
\begin{itemize}
    \item[(i)] $\mathcal{S}_n$ is closed under finite intersections, finite unions, and complements;
     \item[(ii)] $\cS$ is closed under Cartesian products: $A \times B \in \cS_{n+m}$ if $A \in \cS_n$, $B \in \cS_m$; 
     \item[(iii)] $\cS$ is closed under linear projections $\pi:\bbR^{n+1} \rightarrow \bbR^n$: $\pi(A)\in \cS_n$ if $A \in \cS_{n+1}$;
     \item[(iv)] $\cS_n$ contains the zero-sets of all polynomials in $n$ real variables. 
\end{itemize}
Note that the zero-sets of 
polynomials in $n$ variables are the algebraic 
sets of $\bbR^n$.
A structure becomes an \textit{o-minimal structure} when implementing the following tameness constraint: 
\beq 
\boxed{\text{ (v) Definable sets $\cS_1$ of $\bbR$ are unions of finitely many points and intervals. }}
\nonumber
\eeq
We note that the intervals can be closed or open and of finite or infinite length. Remarkably, this seemingly simple condition has many strong implications that justify the notion of having a tame geometry. Its importance for sets of $\bbR^n$ becomes clear if one recalls the projection axiom (iii) of any structure. Eventually all one-dimensional linear projections of a definable set of $\bbR^n$ have to reduce to a finite set of intervals and points.  

Having introduced the sets underlying the tame topology we next need to specify what we mean by a tame map. This is done by requiring that such a map $f:A\rightarrow B$
between two $\cS$-definable sets has a graph that is also $\cS$-definable. We will 
call such maps $\cS$-definable or simply definable for short. We note that the image and preimage of a definable set under a definable map is definable and that  
the composition of two definable maps is definable. Definable maps can be used to 
define definable topological spaces and definable manifolds by requiring the existence of a finite atlas of definable sets with definable transition functions. This notion of definability can equally be applied to the complex geometry using $\bbC^n \cong \bbR^{2n}$. 
There are several general 
results on tame complex geometry and we refer to \cite{PS} for more details and further references. 

It is interesting to point out that o-minimal structures were first introduced in the field of mathematical logic and are part of model theory. A structure represents  sets of `formulas
' and the set-theoretic operations, such as forming intersections, unions, and complements, correspond to 
the logical operations, such as `and' $\land$, `or' $\lor$, and `not' $\lnot$. The important projection property of a structure hereby corresponds to the logical quantifier `exists' $\exists$. 
The introduction of o-minimality as a tameness principle originated from the desire to
make precise statements about 
theories that involve non-trivial functions such as the exponential function \cite{key762106m}.
Historically, the natural tameness property first appeared to be decidability. This would label structures that are built from the natural numbers as being wild, since they admit undecidable statements by the famous G\"odel incompleteness theorems. However, even as of today it is unknown if the structure built from the algebraic sets over the reals when also using the exponential function is decidable. 
O-minimality turned out to be a more tractable tameness criterion. It also excludes considering the set of natural numbers, but allows for proving the o-minimality of many structures that are generated from non-trivial functions as we will explain in more detail below.

\subsubsection*{Cell decomposition theorems}

Among the most useful results in o-minimal structures are the monotonicity theorem and the cell decomposition theorem. The former describes how tame functions $f:(a,b) \rightarrow \bbR$ will look like on any open interval $(a,b)\subset \bbR$. It states that $(a,b)$ can be split into finitely many open intervals and points such that $f$ is either constant or strictly monotonic and continuous on the intervals. 
To obtain a result in higher dimensions, one applies the cell decomposition theorem (see, e.g., ref.~\cite{VdDbook}). A cell decomposition of $\bbR^n$ is essentially a slicing $\bbR^n$ into finitely many pieces, so-called cells, by using definable functions that are also continuous. The cells do not need to be of the same dimension, but rather one starts iteratively building up from lower-dimensional cells to get higher-dimensional cells. The cell decomposition 
theorem now states that one can always find such a decomposition such that any definable set is a finite union of cells. The direct generalization of the monotonicity theorem is the statement that 
for any definable function $f:A \rightarrow \bbR$ one can find a partitioning of $A$ 
into cells, such that it is continuous on each cell.

Note that both the monotonicity theorem and the cell decomposition theorem can be generalized by replacing `continuous' with being $p$ times differentiable (e.g.~being in $C^p$). This might then require a further split of the space into smaller cells, but this can be done while keeping the finiteness of the decomposition intact. However, it is important to stress that while there exists a $C^p$-cell decomposition 
theorem, one cannot generally consider $C^\infty$ or analytic functions. 
In fact, it was shown in \cite{MR1992825} that there are o-minimal structures that define functions that are nowhere analytic on the real line. Nevertheless, for the o-minimal structures prominently used in this work, most notably~$\bbR_{\rm an,exp}$, a $C^\infty$ and analytic cell decomposition does exist (see e.g.~\cite{LionSpeissegger}).

\subsubsection*{Examples of o-minimal structures}

There are many known o-minimal structures and we will introduce the most important examples for us in the following. The smallest structure is denoted by  $\bbR_{\rm alg}$ and is 
generated by the algebraic sets and the requirement for the structure to be closed under the above-mentioned operations. The algebraic sets are the zero-sets of polynomial equations $P(x_1,...,x_n)=0$ in $\bbR^n$ and one shows that also all sets satisfying 
polynomial inequalities $P(x_1,...,x_n)>0$ are then part of the structure. 

It is a non-trivial task to find extensions of 
$\bbR_{\rm alg}$ while preserving the tameness property (v). The for us most relevant extension is the
o-minimal structure denoted by $\bbR_{\rm an,exp}$. To define this structure, one considers more general sets obtained 
by equations of the form 
\beq \label{gen_conP}
P_k(x_1,...,x_m,f_1(x),...,f_m(x))=0\ ,
\eeq 
where $f_i(x)\equiv f(x_1,...,x_m)$ are real-valued functions and $k$ runs over a finite index set. Starting from the sets \eqref{gen_conP} one can generate a structure by including all additional sets that are required to satisfy the defining axioms.\footnote{In particular, one has to include all linear projections of the sets \eqref{gen_conP}. To specify the resulting sets one then has to use also inequalities.}
To guarantee that the structure is o-minimal, i.e.~satisfies the tameness condition (v), the task is to find classes of `sufficiently tame' functions. 
Remarkably, it was shown in \cite{vdDMiller} that the following 
two classes of functions can be used to define an o-minimal structure $\bbR_{\rm an,exp}$:
\begin{itemize}
\item[exp:] Using the real exponential exp$: \bbR \rightarrow \bbR$ as a choice for $f_i$. That this transcendental function can preserve o-minimality was shown in the influential theorem \cite{Wilkie96}.

\item[an:] Using all restricted real analytic functions as choices for $f_i$. Such functions are all restrictions $f|_{B(R)}$ of functions $f$
that are real analytic on a ball $B(R')$ of finite radius $R'$ to a ball $B(R)$ of strictly smaller radius $R<R'$. 
\end{itemize} 
It turns out that also $\bbR_{\rm exp}$ and $\bbR_{\rm an}$, which either use only the exponential or only the restricted analytic functions to extend $\bbR_{\rm alg}$, are o-minimal.

Note that $\bbR_{\rm an,exp}$ has to be used to make the complex exponential $e^z$ definable. To see this, we first note 
that $e^z$ with domain $\bbC$ is never definable, since $e^{z} = e^{r + i \phi} = e^{r}(\cos \phi + i \sin \phi)$ and the graph of the sine- and cosine-functions on 
all of $\bbR$, cannot be definable due to the fact that the projection to the $\phi$-axis results in an infinite discrete 
set of zeros.  To make $e^z$ definable, we restrict the domain of $z$, say by demanding $0\leq \phi \leq a$. Note that $\cos(\phi), \sin(\phi)$ restricted to the domain $0\leq \phi \leq a$ are restricted analytic functions. $e^r$, $r \in \bbR$ is not restricted analytic and $\bbR_{\rm an, exp}$ has to be used to make $e^z$ definable on the domain $0\leq \phi \leq a$. We stress that even 
though $\bbR_{\rm an, exp}$ is significantly larger than $\bbR_{\rm alg}$, there are commonly appearing functions that 
are \textit{not} definable in this structure. 
Most notably, neither the Gamma-function $\Gamma(x)$ on $(0,\infty)$ nor the Riemann Zeta-function on $(1,\infty)$ are definable in $\bbR_{\rm an, exp}$ as shown in \cite{vandendries_macintyre_marker_1997}. 

It is important to stress that the theory of o-minimal structures is 
very rich and still under investigation. To give an example of this, let us note 
that a longstanding question of whether or not one can construct an o-minimal structure that makes both $\Gamma(x)|_{(0,\infty)}$ and $\zeta(x)|_{(1,\infty)}$ definable was only answered 
very recently. 
It has been proved more than two decades ago in \cite{VANDENDRIES200161} that one construct two different o-minimal structures making either one or the other function definable. To show that there is a structure in which \textit{both} are definable was only achieved earlier this year in
\cite{Rolinetal}. This example indicates that, in general, it is not possible to simply combine o-minimal structures to find larger structures. This fact has been known for a longer time already, with the first examples given in \cite{MR1992825}. Whether or not there is a single o-minimal structure that suffices for all physical applications is an open and challenging question. In this work, it will be often sufficient to consider the o-minimal structure $\bbR_{\rm an, exp}$, but it is important to keep in mind that the generalized finiteness properties and the logical completeness statements are present in all o-minimal structures.

\section{Tameness of perturbative amplitudes} \label{sec:tameness_theorem}

In this section we will make our main statement and sketch the proof of the Theorem stated in \S \ref{summary}. For this we first carefully define our setup and the involved spaces. Our starting point is a quantum field theory on a  $d$-dimensional space-time 
defined by a Lagrangian $\mathcal{L}$. We require this theory to be local and 
describe the dynamics of finitely many fields which are either scalars, fermions, gauge fields, or higher form fields. 
We stress that the considered Lagrangians are thus assumed to have only finitely many terms that depend polynomially 
on the fields of the theory. This will ensure that in perturbation theory the physical amplitudes can be computed to a certain fixed loop-level using a finite number of propagators and interaction vertices. 
Our main statement will also need renormalizablity of the QFT, since it relies on the presence of only \textit{finitely} many counterterms. We will comment on the treatment of non-renormalizable effective theories and later return to these cases in \S \ref{tame-EFTs}.

\subsection{Definability Statement} \label{sec:definability_statement}

Let us begin with introducing the precise definability statement. In order to do that we  
denote by $\mathcal{A}_\ell(p,m)$ the considered physical $\ell$-loop Feynman amplitude with $p = (p_1,...,p_n)$ being the $n$ independent momenta of the external states and $m = (m_1,...m_p)$ being the bare 
masses of of the fields of the theory. 
Depending on the dimension $d$ of our theory, it might be necessary to evaluate $\mathcal{A}_\ell(p,m)$
with a renormalization scheme. We will do that using dimensional regularization and denote by  
$\epsilon$ the parameter labelling the dimension $d+\epsilon$ in which the regularized amplitude is 
evaluated. The physical amplitude $\mathcal{A}_\ell$ is then obtained in the $\epsilon \rightarrow 0$ limit.  
It should be viewed as a real map and takes values in the interval $[0,1]$, i.e.~we have a map
\beq \label{cA-map}
   \mathcal{A}_\ell : \mathfrak{M} \times \mathfrak{P}    \rightarrow [0,1]\ , 
\eeq 
where $\mathfrak{M}$ is the momentum space spanned by $p$ and $ \mathfrak{P} $ is the parameter space of the considered quantum field theory and is spanned by the masses $m$ and interaction strengths $\lambda$ extracted from $\mathcal{L}$. 
The general statement that we show in the following is:
\begin{center}
\boxed{\quad \begin{minipage}{13cm}\vspace*{.15cm} The $\ell$-loop amplitude $\mathcal{A}_\ell$ as a map $\mathfrak{M} \times \mathfrak{P}  $ to $\bbR$ is definable in the o-minimal structure $\mathbb{R}_{\rm an,exp}$.\vspace*{.15cm} \end{minipage}\quad }
\end{center}
\noindent
We will show this statement in three main steps: (1) introduce a definable structures on the domain and the target of the amplitude; (2) show that the amplitudes are given by relative period integrals, and (3) use  recent theorems proving the definability of period integrals. 

\subsection{Outline of proof }

Let us now sketch in more detail how the definability of amplitudes can be shown. To begin with, note that 
in perturbative quantum field theory, the $\ell$-loop amplitude $\mathcal{A}_\ell$ is split up in the contribution of Feynman diagrams
\begin{equation} \label{cAgen}
    \mathcal{A}_\ell = \Big| \sum_j f_{\ell,j}(\lambda) I_{\ell,j}(p,m)\Big|^2\, ,\qquad j=1,\ldots, N_{\text{graphs},\ell}\;,
\end{equation}
where $N_{\text{graphs},\ell}$ denotes the number of Feynman diagrams at loop-level $\ell$. 
In this expression we have split off the dependence on the couplings $\lambda$ via functions $f_{\ell,j}$. 
Each $f_{\ell,j}$ is simply a monomial in the various $\lambda$ associated to the appearing vertices.  Therefore, tameness of $\mathcal{A}_\ell$ in $\lambda$ is trivially guaranteed, since \eqref{cAgen} is a sum of finitely many monomial terms. In contrast, the integrals $I_{\ell,j}$ are, 
in general, very complicated functions of the external momenta and possibly the masses of all fields of the theory. Most of the machinery we are going to use is based on integrals only including scalars. But it is always possible to reduce a tensor integral to pure scalar integrals, albeit with different powers of the propagators and in different integer dimensions \cite{Tarasov:1997kx}. We will assume that this procedure has been carried out and all $I_{\ell,j}$ are scalar integrals. We refer the 
reader to \cite{Tarasov:1997kx} for a more detailed discussion of this reduction. 

The physical amplitude is always finite and the divergences cancel separately at any loop level $\mathcal{A}_\ell$. But a single Feynman diagram $I_{\ell,i}$ can have an infinite result. To extract the physical relevant finite piece one expands the Feynman integrals in a Laurent series around the dimension $d+\epsilon$:
\beq 
\label{defIlj}
  I_{\ell,j}(p_1,...p_n,\epsilon) = \sum_{i\ge i_{\rm min}} \epsilon^i I_{\ell,j}^{(i)}\ . 
\eeq 
As the final expression for the amplitude is finite and we are only interested in the amplitude in the dimension $d$, i.e.~the limit $\epsilon\rightarrow 0$, the amplitude is expressible in terms of the coefficients of this Laurent series. The amplitude is the absolute square of the Feynman integrals, thus there is an upper limit on the order of the expansion which will contribute to the finite piece given by $i_{\rm min}$. Thus the amplitude can be written as
\begin{equation}
    \mathcal{A}_\ell= \sum_{j_1=1}^{N_{\text{graphs},\ell}}\sum_{j_2=1}^{N_{\text{graphs},\ell}}\sum_{i=0}^{i_{\rm min}} I_{\ell,j_1}^{(i)}I_{\ell,j_2}^{(i_{\rm min}-i)}\;,
\end{equation}
where we have suppressed the dependence on the couplings $\lambda$.
From now on we will focus on a single object $I_{l,j}^{(i)}$, which we will simply denote by $I$ to avoid cluttering of the notation. We will see that these integrals are definable in $\mathbb{R}_{\rm an,exp}$ and therefore also the amplitude.

We now sketch the a direct argument why the amplitudes are periods of a geometric origin and thus are definable. Therefore, we associate an auxiliary complex manifold $Y_{\rm graph}$ to each Feynman diagram. We denote the complex dimension of $Y_{\rm graph}$ by $d_{\rm graph}$. The details of this construction are rather technical and we postpone their discussion to \S \ref{FeynmanPeriods}. The key point is that 
$Y_{\rm graph}$ admits a moduli space $\cM_{\rm graph}$ of complex structure deformations, i.e.~$Y_{\rm graph}$ 
actually should be thought of as a family of complex manifolds varying over $\cM_{\rm graph}$. The local 
coordinates $z_i$ on $\cM_{\rm graph}$ can be explicitly constructed as polynomials of the external momenta $p$ and masses $m$. The upshot of this construction is that we replace the information $(p,m)$ in the $\ell$-loop integral with complex variables $z_i$ in a definable way by a definable map 
\beq \label{map-to-modulispace}
    \mathfrak{M} \times \mathfrak{P}    \rightarrow  \cM_{\rm graph}\ , \quad (p,m) \mapsto z \ .
\eeq
Henceforth we work on the moduli space $\cM_{\rm graph}$. 
The Feynman integrals are lifted to functions on the moduli space $\cM_{\rm graph}$ and given by 
volumes of cycles of real dimension $d_{\rm graph}$ in the auxiliary geometry $Y_{\rm graph}$. We will discuss this correspondence in detail in \S \ref{FeynmanPeriods}. Concretely, we will recall below that the lifted amplitudes can be written as 
\beq \label{I(z)int}
    I (z) = c^i \int_{\gamma^i} \Omega \ ,  
\eeq
where $\Omega$ is a $(d_{\rm graph},0)$-form changing holomorphically over $\cM_{\rm graph}$ and $\gamma_i$ are $d_{\rm graph}$-chains.
This identifies $I(z)$ as a certain complex linear combination of 
\textit{period integrals}  $\int_{\gamma^i} \Omega$. If one also allows for boundaries of the integration domain, they are known as relative periods.

This type of argument is not new. It is well known that Feynman integrals can be interpreted in terms of GKZ systems. Many examples have been worked out explicitly e.g.~in \cite{Boos1991,Fleischer:2003rm,Feng:2019bdx}. 
The upshot of this is that any scalar Feynman integral can be represented as a linear combination of solutions to GKZ systems.\footnote{ This correspondence requires that the Newton polytope assigned to the Feynman graph is full dimensional. This is always the case except for tadpole graphs. These have to be canceled by adding counterterms in the renormalization procedure and we will thus assume they are absent.}

As a final step in our definability argument we will use a remarkable result due to Bakker and Mullane \cite{bakker2022definable} that ensures the definability of the period integrals. Concretely, the corollary 1.3 of \cite{bakker2022definable}, roughly states that
\begin{center}
\boxed{\quad \begin{minipage}{13cm}\vspace*{.15cm} Relative period integrals are definable in the o-minimal structure $\mathbb{R}_{\rm an,exp}$.\vspace*{.10cm} \end{minipage}\quad }
\end{center}
\noindent
Note that the statement is general if one considers integrals over rational algebraic forms defined on a family of smooth algebraic varieties. If the variety is not smooth, additional steps in the argument are needed, as discussed at the end of section \ref{FeynmanPeriods} and in section \ref{divergences_reg}.
The definability result of \cite{bakker2022definable} is intimately related to the definability of the general period maps \cite{bakker2020tame,bakker2020definability} and turns out to be most directly applicable in our context.
To use this result, we have thus to show that the integrals in \eqref{I(z)int} are actual period integrals. This will be studied in the next section in detail.\footnote{Note that our reasoning is similar to \cite{Bogner:2007mn} where it was shown that the $I_{l,j}^{(i)}$ are periods in the sense of Kontsevich and Zagier \cite{Kontsevich2001}. The arguments for definability require to 
extend these to periods of families varying over a complex moduli space}

\subsection{Feynman Integrals, Periods, and Definability}
\label{FeynmanPeriods}

Let us proceed to the detailed proof.
We start with a review of some representations of Feynman integrals, the Symanzik and Lee-Pomeranski parameterizations. We then describe how the Feynman integrals are obtained as linear combinations of period integrals. Finally, we argue for the definability of Feynman integrals using the definability of the period map.

Let us consider again an $\ell$-loop amplitude $\mathcal{A}_\ell$
derived in a local quantum field theory as in \S \ref{sec:definability_statement}. 
This amplitude is a map depending on the independent external momenta and the masses as in \eqref{cA-map}. In practice, the amplitude is derived from a finite sum of Feynman diagrams with associated Feynman integrals. In the case of a pure scalar $\ell$ loop integral in $d$ dimensions the integral takes the form
\begin{equation}
\label{integral1}
    I(p,m)=\int\left(\prod_{r=1}^L \frac{\mathrm{d}^dk}{i \pi^{d/2}}\right)\left(\prod_{j=1}^n\frac{1}{D_j^{v_j}}\right)\;,
\end{equation}
where $D_j=p^2_j-m^2_j$ are the propagators\footnote{The propagators are understood with a suitable contour deformation around the poles.} of the theory and the $v_j\in \mathbb{Z}$ the exponents of the propagators. In scalar theories one considers $v_j=1$, but we keep $v_j$ general in order to also be able to treat non-scalar fields. We assume that in 
case the amplitude requires to include a tensor structure,  e.g.~arising from gauge fields,
that a reduction to scalar integrals has been performed. In such a case $d$ might not be 
the actual space-time dimension, but a dimension fixed in the reduction. 
At each loop level there are only a finite number of Feynman integrals. 

To make contact with the geometric interpretation it is useful to rewrite the integral in different representations. A standard trick in the computation of Feynman integrals is to replace products of propagators with a single sum at the cost of introducing Schwinger parameters $x_i \in \bbR$. I.e.~one uses the identity
\begin{equation}
    \prod_{j=1}^n \frac{1}{D_j^{v_j}}=\frac{\Gamma\left(v\right)}{\prod_{j=1}^n\Gamma(v_j)}\int_{x_j\geq 0}\mathrm{d}^n x\;\delta\Big(1-\sum_{j=1}^n x_j\Big)\, \frac{ \prod_{j=1}^n x_j^{v_j-1} }{ \sum_{j=1}^n x_jD_j}
\end{equation}
to replace the propagators in \eqref{integral1}. Here we have defined the sum of the propagator exponents
$v=\sum_{j=1}^n v_j$ to shorten the expressions. One can then perform the integrals over the loop momenta and arrives at the following representation \cite{Lee:2013hzt}:
\begin{equation}
\label{integral2}
   I=\frac{\Gamma\left(v-\ell d/2\right)}{\prod_{j=1}^n\Gamma(v_j)}\int_{x_j\geq 0}\prod_{j=1}^n\mathrm{d} x_j x_j^{v_j-1}\;\delta \Big(1-\sum_{j=1}^n x_j\Big )\, \frac{F^{\ell d/2-v}}{U^{(L+1)d/2-v}}\;.
\end{equation}
The $F=F(x,p,m)$ and $U=U(x,p,m)$ in this expressions are so-called Symanzik polynomials, which are homogeneous polynomials of degrees $\ell$+1 and $\ell$ in the Schwinger parameters. Their exact form can be determined algorithmically from the Feynman graph using graph theoretic methods \cite{Bogner:2010kv}. The details are described in appendix \ref{appendix1}.

This ratio of polynomials is still not perfectly suited for a geometric interpretation, for which one would prefer to have only a single polynomial. There are two observations which help with this problem. First, in some cases the representation \eqref{integral2} simplifies. $\ell$-loop banana integrals have $\ell$+1 propagators, thus in two dimensions one has
\begin{equation}
    \frac{F^{\ell d/2-v}}{U^{(\ell+1)d/2-v}}=\frac{1}{F}\;,
\end{equation}
i.e.~the exponent of the second Symanzik polynomial $U$ vanishes. Second, for the general case it was observed in \cite{Lee:2013hzt} that it is always possible to rewrite the representation \eqref{integral2} in terms of a single polynomial 
\beq \label{def-G}
    G=U+F\ .
\eeq 
The representation of the Feynman integral obtained this way is named Lee-Pomeransky representation after the authors of \cite{Lee:2013hzt}. The final representation is then
 \begin{equation}
 \label{feynmanintegral}
     I=\frac{\Gamma\big(\frac{d}{2}\big)}{\Gamma\Big(\frac{(\ell+1)d}{2}-v\Big)\prod_{j=1}^n\Gamma(v_j)}\int_{x_j\ge 0}\prod_{j=1}^n\mathrm{d} x_j x_j^{v_j-1} G^{-\frac{d}{2}}\;.
 \end{equation}
 The equivalence of \eqref{feynmanintegral} and \eqref{integral2} can be seen by inserting a 1 into \eqref{feynmanintegral}  in the form of
\begin{equation}
    1=\int \mathrm{d}s\, \delta\Big(s-\sum_{j=1}^n x_j\Big)\;.
\end{equation}
After rescaling the Schwinger parameters as $x_j\rightarrow s x_j$ and performing the $s$ integral one arrives at the representation \eqref{integral2}, which shows the equivalence.

Let us now describe how the Lee-Pomeransky representation can be used to realize the scalar Feynman amplitude as a period integral in an auxiliary complex algebraic variety $Y_{\rm graph}$ associated to the considered graph. We begin by  
viewing $x_i$ as \textit{complex} coordinates of a  complex weighted projective spaces.  In even dimensions\footnote{The even dimension is necessary to avoid a square root in the defining polynomial. Later we will describe how to deal with $d\in\mathbb{R}$ for dimensional regularization, which also allows for odd values.} $Y_{\rm graph}$ is then defined as the hypersurface 
\beq \label{hypersurface}
P(x_i) \equiv G(x_i)^{d/2} = 0\ , 
\eeq 
where $G$ is the homogeneous polynomial arising in the 
 Lee-Pomeransky representation introduced in \eqref{def-G}. The hypersurface \eqref{hypersurface} can be viewed as a special case of a general hypersurface $P(a_j,x_i)=0$ specified by the scaling weights of the $x_i$ and depending on complex parameters $a_j$
 that arise as coefficients of the individual monomials in $x_i$. It turns out that different choices for the $a_j$
 correspond to different choices of a complex structure 
 on $Y_{\rm graph}$. The physical parameter space is then a slice in the space of all complex parameters $a_j$. 
 
 It is important to note that $Y_{\rm graph}$ can be a singular manifold.  However, as  described in \cite{Hironaka}, the singularities can be removed by performing consecutive blow-ups till a smooth space is reached. This can be an involved procedure in practice (see e.g.~\cite{weinzierl_feynman_2022} for details and more references), but the existence of such a resolution suffices for our arguments. 
 Having found a smooth hypersurface, we need to recover the Lee-Pomeransky polynomial $G$ introduced in \eqref{def-G}. In order to do that the coefficients $a_j$ are evaluated in terms of the independent external momenta and masses by simple algebraic expressions (see appendix \ref{appendix1} for details). In the geometric setting we are therefore have $(p,m)$ to set a choice of complex structure on $Y_{\rm graph}$. In general, also the $a_j$ parameterize the complex structure in a redundant way. The independent choices are obtained by appropriately combining the $a_j$ into fractions invariant under the reparameterization symmetries of $P$. The resulting degrees of freedom are the complex structure deformations $z^I$ of the hypersurface \eqref{hypersurface} and can be shown to span
 a moduli space $\cM_{\rm graph}$. This construction 
 thus provides us with a map  between the momenta and 
 masses and the complex structure moduli of a smooth hypersurface as already 
 mentioned in \eqref{map-to-modulispace}.

The construction of $Y_{\rm graph}$ is motivated by the 
fact that there is a natural class of integrals that can be determined on such a variety. Denoting by $\gamma_i$ 
a $d_{\rm graph}$-dimensional chain, as in \S \ref{sec:definability_statement}, we introduce the period integral 
\begin{equation}\label{periodintegral}
    \omega_i=  \int_{\gamma_i}\Omega \equiv \int_{\gamma_i} \frac{\mathrm{d}x_1\wedge\mathrm{d}x_2\wedge\ldots\wedge\mathrm{d}x_n}{ P(a_j,x_i)}\;,
\end{equation}
where we have introduce the ($d_{\rm graph}$,0)-form $\Omega$.
The key point of this construction is that, apart from the different integration domain, the integrals \eqref{feynmanintegral} and \eqref{periodintegral} are of the same form when evaluating $P(a_j,x_i)$ for the values $a_j$ needed in identifying the Lee-Pomeransky polynomial $G$ with $P$. If $\gamma_i$ would be a closed chain, i.e.~a cycle, the integral in \eqref{periodintegral} would be a linear combination of pure periods. But as $\gamma_i$ is an open chain, it can only be expressed in terms of a combination of pure periods and a relative period.%
\footnote{The relative periods are elements of the relative cohomology 
$H^{\bullet}(\widetilde{\mathbb{P}^{n-1}}\setminus {\widetilde{Y_{\rm{graph}}}},\widetilde{B}\setminus \widetilde{B}\cap \widetilde{Y_{\rm{graph}}})$, 
where the $\sim$ denotes appropriate blow ups and $B$ is the divisor corresponding to $x_1x_2\cdots x_n=0$.}
These types of integrals are definable by corollary~1.3 of \cite{bakker2022definable}. 
As long as the integral in \eqref{periodintegral} is finite that is the end of the story. But in some cases divergences can appear, which have to be regularized. The usual approach to this problem is dimensional regularisation, where the dimension $d$ is slightly shifted to $d+\epsilon$. But as the dimension appears in the exponent of the defining polynomial \eqref{hypersurface} this appears to destroy the direct correspondence to the geometry. But this is remedied by the GKZ property of the integrals. This will be discussed in the next section.

\subsection{Divergences and Regularization} \label{divergences_reg}
To extract physical information out of divergent diagrams it is necessary to regularize them first. The effects of this regularization are nicely understood as deformations of a so-called Gel’fand-Kapranov-Zelevinsky (GKZ) system of differential equations. Period integrals are an examples of such systems. The Feynman integrals fulfill the same system of differential equations, which was proven in \cite{delaCruz:2019skx} using the Lee-Pomeransky representation. 
The data of the GKZ system is encoded in two objects, the configuration matrix $A$, which contains the exponents of the polynomial $P$, as well as a vector $v$ which encodes the dimension and the powers of the propagators:
\begin{equation}
    v=\big\{\tfrac{d}{2},v_1,v_2,\ldots,v_N\big\}\;.
\end{equation}
Divergences can appear for integer values of these parameters. To remove these, dimensional regularization replaces $d\rightarrow d-2\epsilon$, while analytic regularization replaces $v_i\rightarrow v_i-\tilde{\epsilon}$. Here we will only focus on dimensional regularization, but the arguments for analytic regularization are equivalent. The regularized Feynman integrals $I_{\ell,j}^{(i)}$ as defined in \eqref{defIlj} are then obtained from the solutions of the deformed system as the coefficients of the series expansion in $\epsilon$.

In the integral representation \eqref{periodintegral} this expansion in $\epsilon$ would lead to logarithmic terms, rendering a direct geometric interpretation difficult. But the integral \eqref{periodintegral} is the representation for the fundamental period. To obtain the full set of periods different cycles have to be chosen. A basis of solutions can be encoded into the $\epsilon$-expansion of the fundamental period using the Frobenius method. This expansion and the expansion in the regularization parameter are actually equivalent expansions, which follows from \cite{emad2016}, where it was shown that the D-modules of the GKZ system and the Feynman diagrams agree. This implies that the $\epsilon$ expansion of the periods is the same expansion as the epsilon expansion of the Feynman integral, as both form a basis of the same D-module.

For the amplitude to be finite, the coefficients of the negative powers of $\epsilon$ in the Laurent expansion have to cancel. For arbitrary parameters in the Lagrangian this will generally not be the case and the introduction of counter terms becomes necessary. By the BPHZ theorem \cite{10.1007/BF02392399,cmp/1103815087,cmp/1103841945} it suffices in the case of a renormalizable theory to introduce counterterms to the superficially divergent graphs. There are only a finite number of such graphs and therefore only a finite number of counterterms needed. After introducing these counterterms, the resulting Lagrangian is still tame if the original Lagrangian was tame. The renormalized Lagrangian can then be taken as the starting point and the arguments of the previous section still apply. Therefore the $l$-loop amplitudes in the renormalized theory are definable in $\bbR_{\rm an,exp}$. 

This argument also applies to correlation functions including a finite number of perturbatively irrelevant 
operators (operator dimension $\Delta>d$).  Renormalization of such operators (sometimes called composite operators)
now requires introducing a larger set of counterterms, generally including all lower dimensional operators
\cite{collins1985renormalization}.  However this set is finite and the number of graphs involved is still finite.

Adding irrelevant operators to the action with finite couplings leads to much more severe problems. In this case, an infinite number of counterterms is needed, which threatens the tameness of the Lagrangian. One might hope, however, that this issue can be addressed when having knowledge about the UV completion of the theory. As we will discuss in \S \ref{tame-EFTs}, it is plausible to conjecture that only 
those theories are compatible with quantum gravity that admit a tame Lagrangian. 
Interpreting the non-renormalizable theory as an effective theory, valid up to a cutoff $\Lambda$, which has such a UV completion, this tameness conjecture then implies that the infinite number of counterterms need to combine into a tame Lagrangian. Note that these ideas might require to go beyond the purely perturbative analysis of the theory.  Moreover, we will see in \S \ref{nonpert-challenges} that 
it is easy to write down UV Lagrangians that are not tame, which makes the importance of quantum gravity plausible. Starting from a tame UV theory, the cutoff can be lowered and all fields heavier than the cutoff integrated out. The effects of the RG flow on the tameness of the amplitudes when lowering the cutoff is discussed in more detail in \S  \ref{RG-flow}. In summary we expect the tameness to be preserved, such that the coefficients of an expansion of the amplitudes in the cutoff should be tame functions.


\section{Are non-perturbative QFT results tame?}
\label{np-tameness}

So far we have focused on the perturbative approach to QFT up to a fixed loop-order. While the proof of the tameness in momenta and couplings at finite loop order is very general and requires little more than the general structure of Feynman integrals, it is essentially perturbative. For non-perturbative amplitudes the question of tameness is more subtle. Even simply trying to extend it by summing up Feynman diagrams to all loop orders would result in an infinite sum which is not guaranteed to respect tameness. Still, the hypothesis that a particular amplitude is tame makes perfect sense as a claim about the exact (non-perturbative) theory. The intuition that QFT
becomes simple at weak coupling also suggests that it could be true in some generality, as do arguments that string compactifications and QFTs which can be coupled to a (hypothetical) other quantum gravity theory are finite in number. Indeed, in many examples it turns out that the full non-perturbative partition functions and therefore the amplitudes are tame at least in the couplings of the theory. 
In this section we will gather some evidence for the tameness of non-perturbative QFTs and study possible challenges. In particular we will see that non-perturbative tameness is closely related to the famous no global symmetry conjecture.

\subsection{Partition functions}
\label{sec:partition}
In this section we discuss some examples of exactly solvable theories for which the full partition functions including all non-perturbative terms can be computed and shown to be tame.

\subsubsection*{0d QFT: Sine-Gordon model}

As our first example we consider the sine-Gordon model in zero dimensions, i.e.~we study the theory on a point. This model has a potential $V = 2 \lambda \sin^2(\phi)$, where $\lambda$ is a coupling constant. As we are working in zero dimensions the field $\phi$ is simply a real number. The path integral defining the partition function of this theory reduces to the standard integral \footnote{We choose the coefficient of the coupling such that the normalized partition function $\mathcal{Z}/\mathcal{Z}_0$ becomes exactly a geometric period, see equation \eqref{periodrepresentation}. This is purely for aesthetics and any rescaling $\lambda \rightarrow a \lambda$ would work.}
\begin{equation} \label{eq:sgmodel}
    Z=\int_{-\pi}^{\pi}d\phi\, e^{4 \lambda \sin^2(\phi)}  =2\pi e^{2\lambda}I_0(2  \lambda)\,.
\end{equation}
Here $I_0(x)$ is the modified Bessel function of the first kind. For this function we can find an explicit geometric realization. To see this, one constructs a gauged linear sigma model (GLSM)  corresponding to the charge vector
$l=\{-2,1,1,1\}$. The geometry described by this model has the fundamental period
\begin{equation}
\label{periodrepresentation}
    \omega_0=\sum \frac{x^n \Gamma (2 n+1)}{\Gamma (n+1)^3}=e^{2 x} I_0(2 x)\;,
\end{equation}
which is exactly the partition function of the Sine-Gordon model. Note that the sum of the charges does not cancel and the GLSM thus does not describe a flat space. Nevertheless, this connection shows that $e^{2 x} I_0(2 x)$ is a period integral and thus $\bbR_{\rm an,exp}$-definable by \cite{bakker2022definable}. This implies that the partition function of the 0d Sine-Gordon model as a function of $\lambda$ is definable in $\bbR_{\rm an,exp}$.
However we will see in \S \ref{nonpert-challenges} that the generalization to other 0d models is an open question.

\subsubsection*{1d QFT}

Let us consider the general finite temperature partition function: given an energy spectrum $E_n$,
possibly depending on other couplings (schematically denoted $\vec \lambda$), this is
\begin{equation}
    Z(\beta,\vec\lambda) = \sum_n \exp\big[ -\beta E_n(\vec\lambda) \big] = \int dx \, G(x,x;\beta)
\end{equation}
where $G$ is the Euclidean time propagator. We first consider the harmonic oscillator with $V(x) = \frac{m^2}{2} x^2$ 
, then
\begin{equation}
    Z(\beta,m) = \frac{ 1 }{ 2 \sinh  \beta/(2m) } 
\end{equation}
is definable in $\mathbb{R}_{\rm an,exp}$ for $\beta,m\in (0,\infty)$. 

What about more general potentials?
One might argue on physical grounds that the energy levels and partition function will be tame under
variations which preserve the large field behavior.
One can show \cite{simon1970coupling,kato2013perturbation} that starting from a potential 
with a discrete and nondegenerate spectrum, and adding a relatively bounded perturbation (so, preserving the
large field behavior),
the energy levels and partition function are analytic in an open region containing this starting point.
Then, one type of nonanalyticity which can appear is the branch points associated with an eigenvalue degeneracy,
as one can see for finite matrices.  These are also controllable and (if the degeneracies are finite)
are still consistent with tameness.  This leaves singular perturbations (changing the large field
behavior) for which the situation is not at all obvious.

Checking that these partition functions are definable in an o-minimal structure is quite nontrivial. 
Already the anharmonic oscillator with $V(x) = m^2 x^2 + \lambda x^4$ does not seem to have a solution in elementary functions, even for the energy levels. One way to approach this problem is to consider the spectrum in the WKB approximation.  Allowed energies must satisfy the Bohr-Sommerfeld quantization condition,
\begin{equation} \label{eq:wkb}
    S  = \oint p dx = 2 \int dx\, \sqrt{ 2(E_n - V(x)) }= 2\pi (n+1/2) \hbar \, .
\end{equation}
Note that this is a period on the Riemann surface $E=p^2/2 + V(x)$ where $p,x$ are complexified.
It is then tempting to speculate that the energies in this approximation are 
tame functions of the parameters in the potential and that this also 
applies to the associated finite temperature partition function. 

\subsubsection*{2d QFT: Linear sigma models from string compactifications}
\label{twodqft}

As our next example we consider gauged linear sigma models (GLSM) that are two-dimensional $\mathcal{N}=(2,2)$ supersymmetric field theories.  We are interested in the cases in 
which these flow in the infrared to the non-linear sigma model of a type II string on a compact Calabi-Yau threefold. 
In this situation the sphere partition function of the GLSM can be expressed in terms of the K\"ahler potential as \cite{Gomis:2012wy,Jockers:2012dk}: 
\begin{equation}
    Z_{S_2}=\exp(-K)=\ov{\Pi}\Sigma\Pi\;,
\end{equation}
where $\Pi$ are the periods in an integral symplectic basis and $\Sigma$ is the symplectic pairing. The periods are definable as functions of the moduli, which are identified with the Fayet-Iliopoulos parameters and $\theta$-angles of the GLSM. Using the definability of the period integrals we thus show that the sphere partition function is definable in $\bbR_{\rm an,exp}$ as a function in these parameters.  This is also the case for disk partition functions, which compute the central charges of 
Dirichlet branes \cite{Ooguri_1996}.
Note that the partition function also depends on the charges of 
the multiplets in the GLSM. These are discrete parameters which implies that tameness in these parameters requires that the inequivalent choices are only taken from a finite set. This matches nicely with the conjecture that there are only finitely many inequivalent compact Calabi-Yau threefolds.  

\subsubsection*{2d QFT: Free Yang-Mills theory}

As an example of a two-dimensional theory we take the free Yang-Mills theories. In two dimensions these have no perturbative degrees of freedom, but the theories still include non-perturbative effects. The partition functions for a $SU(N)$ group were computed in \cite{Witten:1991we} with the result

\begin{equation}
    {Z}=\sum_{R}\mathrm{dim}(R)^{\chi}e^{-\frac{\lambda A}{2N} C_2(R)}\,,
\end{equation}
where the sum runs over the irreducible representations $R$ of the gauge group. In this expression we denoted by $A$ and $\chi$ the area and Euler characteristic of the spacetime, respectively. $C_2$ is the quadratic Casimir of the gauge group and $\lambda=g^2N$ is the 't Hooft coupling.
As an example we take the $SU(2)$ partition function on a torus. For this theory the partition function becomes
\begin{equation}
   {Z}_{SU(2)}=e^{\frac{A\lambda}{16}}\left(\theta_3(e^{-\frac{A\lambda}{16}})-1\right).
\end{equation}
The definability of theta-functions on their fundamental domain was shown in~\cite{Peterzil2013}. Thus the free $SU(2)$ Yang-Mills theory provides another example of a non-perturbatively definable partition function for all $A, \lambda >0$. Note that this result naturally extends to many 
other settings in which theta-functions specify the partition functions.

\subsubsection*{Non-critical M-theory and 2d strings}

Two-dimensional non-critical Type 0A and 0B string theories have been studied in much detail \cite{Klebanov:1991qa,Douglas:2003up}. These theories admit one free parameter $\mu$ which allows one to define a perturbative expansion. At the non-perturbative level these theories are completed by matrix models. 
We are interested in checking the tameness of the partition function of these two-dimensional string theories in the parameter~$\mu$. 

In \cite{Horava:2005tt} it was shown that the matrix models of two-dimensional non-critical string theory arise as solutions of a three dimensional non-critical M-theory. This M-theory also depends on single free parameter, $\tilde{\mu}=g_{\rm M}^{-2/3}$, which is identified with the free parameter $\mu=g^{-1}_s$ of the string theories. Compactifying this non-critical M-theory on a thermal circle of radius $R$ leads to a theory which is dual to the topological A-model on the conifold \cite{Horava:2005wm}. The partition function 
$Z_{\rm M}(R,g_{\rm M})$ of the non-critical M-theory is equal to the partition function $Z_{\rm A}(t,g_{\rm A})$ of the topological A-model. In this identification $g_{\rm M}$ is mapped to the K\"ahler modulus of the conifold as $t = 2\pi R g^{3/2}_{\rm M}$ and $R$ is mapped to $g_{\rm A} = 2\pi i R$ in the A-model. The A-model partition function takes the 
form  
\begin{align}
    \log Z_{\rm A} =&  \frac{1}{g_{\rm A}^2} \Big(p_{\rm A}(t ) + \frac{t^3}{12} - \text{Li}_3\big(e^{-t}\big) \Big) + \Big(C_{\rm A} - \frac{t }{24} - \frac{1}{12} \log\big( 1- e^{-t}\big) \Big)  \nonumber\\
    & + \sum_{n=2}^{\infty} g_{\rm A}^{2n-2} \left(\frac{B_{2n} B_{2n-2}}{2n (2n-2)(2n-2)!} + \frac{B_{2n}}{2n(2n-2)!} \text{Li}_{3-2n}(e^{-t})\right)\ ,
\end{align}
where $p_{\rm A}$ is a quadratic polynomial, $C_{\rm A}$ is a constant, and $B_{n}$ are the Bernoulli numbers. 
To leading order in $g_{\rm A}$ this is the genus zero prepotential of the conifold. 
Recall that the periods are given by polynomials in $t$ and the derivatives of the prepotential. Using the $\bbR_{\rm an,exp}$-definability of the periods in the K\"ahler modulus $t$, we could infer the definability of the partition function in $\mu$ for this leading term after integration.  
In this example, however, the tameness can also be directly inferred from the fact that the appearing functions Li$_n$ and the exponential function are definable in $\bbR_{\rm an,exp}$ (see appendix \ref{tame-hypergeometry}). Higher orders in the $1/R$-expansion correspond to higher genus corrections in the topological A-model. Due to the $\bbR_{\rm an,exp}$-definability of the appearing functions we infer that tameness in $g_{\rm M},R$ persists at finite genus. For the all genus partition function we are confronted with 
the same problem as encountered in non-perturbative QFT, since the infinite summation could destroy the tameness in the coupling $g_{\rm A}$ of the A-model and therefore in the radius of the M-theory. 
It would be interesting to use the recent insights \cite{Alim:2022oll} to also show tameness in $R$. 

There is an obvious obstruction for tameness in the string theory setting; the existence of an infinite number of fields. These lead to infinitely many poles in amplitudes that when evaluating them as 
a function of external momenta. Clearly, this violates tameness. The situation in two dimensions is slightly different compared to the ten dimensional theories, as there are no transversal directions. Therefore, there are only finitely many perturbative degrees of freedom. Nevertheless, there are still infinitely many so-called discrete states with fixed momenta \cite{Witten:1991zd}, which show up as poles in the amplitudes \cite{DiFrancesco:1991daf}. The simplicity of two dimensional string theory allows to identify the structure behind these states, which form a $w_\infty$ algebra. From the M-theory point of view these arise due to the two dimensional solution of the theory spontaneously breaking parts of the three dimensional diffeomorphism group. The discrete states are then corresponding to the generators of the broken symmetry. The infinite discrete states thus become part of a continuous symmetry and there is no contradiction for amplitudes to be tame. A definite statement would require a better understanding of the amplitudes themselves. Nevertheless, the duality to the topological A model and the tameness of the periods suggests the tameness of these amplitudes.

\subsection{Challenges for tameness in non-perturbative QFT} \label{nonpert-challenges}

In this subsection we will explain some of the challenges that one has to face in order to  establish tameness results at a non-perturbative level. Firstly, we will show that even for simple 
settings, in which we expect tameness to persist at the non-perturbative level, we need new  mathematical definability results going beyond those for period integrals.
For example, we show that the partition function of the zero-dimensional $\phi^4$-theory is 
given by an exponential period, for which definability has not been established.  
Secondly, we discuss the issues that can arise when the UV theory itself contains 
non-tame functions. We argue, in particular, how global symmetries of infinite order 
challenge tameness and how this is linked to some conjectural properties of quantum 
gravity. 

\subsubsection*{Zero-dimensional partition functions and exponential periods}
\label{sss:zero}

Let us consider zero-dimensional $\phi^4$-theory and determine its partition function. 
The action of this theory is given by 
\begin{equation}
    S=\frac{m^2}{2}\phi^2+\frac{\lambda}{4!}\phi^4\;.
\end{equation}
The parameters of the theory are $m$ and $\lambda$. We assume them to be 
non-negative real numbers to ensure that the path integral converges.
In the free field case the partition function is simply a Gaussian integral ${Z}(\lambda=0)=\frac{\sqrt{2\pi}}{m}$ and clearly definable in $m$. In the case of a non-zero value of the coupling the integral can still be performed with the result
\begin{equation} \label{phi4-periods}
    {Z}=\sqrt{\frac{3}{\lambda}}e^{\frac{3m^4}{4\lambda}}\,m\, K_\frac{1}{4}\left(\frac{3m^4}{4\lambda}\right)\;,
\end{equation}
where $K_\frac{1}{4}(x)$ is the modified Bessel function of the second kind. This is a non-oscillating, exponentially decaying function. It can be rewritten in terms of confluent hypergeometric functions $_1F_1$. While this shows some analogy of the partition function \eqref{phi4-periods} with geometric periods, which can involve e.g.~$_2F_1$, there are important differences that we want to discuss momentarily. Before doing this, let us note that $K_{\nu}(x)$ is an analytic function on the real line. This implies that it is restricted analytic for any finite length interval $x \in [x_0,x_1]$ and therefore $K_\frac{1}{4}(x)|_{[x_0,x_1]}$ is definable in $\bbR_{\rm an,exp}$. However, the weak coupling limit $\lambda\rightarrow 0$ in \eqref{phi4-periods} is at $x\rightarrow \infty$ and we would like to have a definability statement including this limit. It turns out that this is an open question, but there is a relatively simple argument at least for this example (see \cite{Grimm:2023xqy} for details).\footnote{It turns out that the functions $K_\alpha$ can be defined in the Pfaffian closure of $\bbR_{\rm alg}$. Pfaffian closures of o-minimal structure preserve tameness, thus the modified Bessel functions are tame functions. We like to thank Lou van den Dries for the proof of the definability in the Pfaffian closure.
\label{footnotevandenDries}}

The modified Bessel function of the second kind can be written as an integral in the following 
way. We recall that 
\begin{equation}
    K_\frac{1}{4}(x) = \frac{\sqrt{\pi} x^\frac{1}{4}}{2^{1/4} \Gamma\big(\frac{3}{4}\big)} \int_1^\infty  
    e^{- x t}  \frac{dt}{(t^2 - 1)^\frac{1}{4}} \ . 
\end{equation}
for $x>0$. Note that this expression involves an integral over an algebraic form $\omega = dt/(t^2-1)^{1/4}$, as for a period integral, but now includes an additional exponential suppression factor $e^{-xt}$. The generalized notion one can introduce to capture these cases are so-called 
exponential periods of the form 
\begin{equation} \label{exponentialPi}
    \Pi_{\rm exp} = \int_{\Sigma} e^{-f} \omega\ , 
\end{equation}
where $f$ is an algebraic function and $\omega$ is an algebraic form. A precise definition was given in \cite{Kontsevich2001}. In this notion one defines $\Pi_{\rm exp}$ to be a special complex number, which can be written as an integral of the form \eqref{exponentialPi} with  $f$, $\omega$, and $\Sigma$ having special properties stated in \cite{Kontsevich2001,CHH}. In \cite{CHH} it was shown that the real and imaginary parts of $\Pi_{\rm exp}$ are volumes of certain definable sets. However, strong theorems, as the ones in \cite{bakker2020tame,bakker2020definability,bakker2022definable} for the period map and period integrals, are still missing. To make progress in this direction, it 
would be interesting to obtain definability results for certain exponential motives defined in the foundational work \cite{Fresan}. This gives a framework to consider $ \Pi_{\rm exp}(x)$ as being obtained from a suitable cohomology that varies over some space parameterized by $x$. It is expected that this gives a framework to discuss, for example, the definability of the modified Bessel functions.\footnote{We would like to thank Bruno Klingler for discussions on this point.} We find it interesting that tameness at the non-perturbative level forces us to 
tackle a new class of functions. 

Clearly, it would be helpful to know whether all the functions $\Pi_{\rm exp}(x)$ are 
definable in an
o-minimal structure replacing $\bbR_{\rm an,exp}$. 
In part II we will define a structure $\bbR_{\rm QFT0}$ to which 0d QFT observables belong. 
Our question will become, is $\bbR_{\rm QFT0}$ o-minimal and if so, 
is it a new structure or simply $\bbR_{\rm an,exp}$.\footnote{This question has partly been settled in version 2 of this work, since it is now clear that $\bbR_{\rm an,exp}$ is too small to define the Bessel function. See footnote \ref{footnotevandenDries}.}

\subsubsection*{One-dimensional partition functions and quantum periods}
\label{sss:one}

There are reasons to think that quantum mechanics also leads to a new class of functions,
possibly requiring a different definition of tameness.
A very interesting approach to the full quantum problem is 
the exact WKB method, see \cite{ito_tba_2019} and references therein.
In a complicated way explained there, the spectrum can be determined using a modified
Bohr-Sommerfeld condition \eqref{eq:wkb} defined in terms of ``quantum periods.''
Another important relation discussed in this literature is to $1+1$ integrable QFT \cite{dorey_anharmonic_1999}.
In part II we will define a structure $\bbR_{\rm QFT1}$
in terms of observables of Euclidean time quantum mechanics, and ask:
is it o-minimal?

\subsubsection*{Counterexamples, global symmetries, and tameness in the UV}

Having discussed situations where we expect tameness to be present, let us now turn 
to cases where tameness is absent and discuss reasons and remedies for this. We can distinguish various classes of counterexamples 
according to the restrictions we place on the
UV definition of the theory.  For example, we might not be surprised to find 
that a theory whose UV Lagrangian includes non-tame functions has 
a non-tame partition function. A priori there 
appears to be nothing wrong to include a non-tame function in the definition in the UV theory, 
but we will argue shortly that tameness might be related 
to the consistency of the theory with quantum gravity. We will return to this issue in \S \ref{tame-EFTs}.

The basic example here is a theory with a theta angle $\theta$, such as 4d QCD.  If we regard the
partition function as a function of $\theta\in\bbR$ then of course it is not tame due to the 
presence of a periodic potential $\cos \theta$.  This issue
is easily remedied by identifying $\theta\cong\theta+2\pi$ and taking the domain to be $\theta\in[0,2\pi)$. More generally, let us consider a theory depending on couplings $\lambda$ varying over some parameter space $\mathfrak{P}$. We want to study the symmetry 
group $G$ acting on $\lambda$, in some faithful representation, such that the 
partition function is invariant 
\begin{equation}
    Z(g\cdot \lambda) = Z(\lambda) \ .
\end{equation}
If $G$ is discrete and admits infinitely many elements that generate a discrete  set of distinct $\lambda$-images in $\mathfrak{P}$, 
we realize that any non-trivial $Z(\lambda)$ cannot be a definable function on $\mathfrak{P}$. This can be remedied by considering $Z(\lambda)$ on the quotient $\mathfrak{P}/G$, which physically means that we consider the symmetry $G$ to be gauged. The restricted $Z(\lambda)$ might then be a tame function. 
This is precisely what happens in the restriction of the cosine to $\theta\in[0,2\pi)$.
More general, in many physical settings, the quotienting by the discrete symmetry is a crucial part of the construction and yields a definable partition function. 
For example, the modular symmetries of the torus 
partition function of a two-dimensional conformal field theory on the string world-sheet are 
gauge symmetries and $Z$ on the fundamental domain is $\bbR_{\rm an,exp}$-definable by \cite{Peterzil2013}. 

It is interesting to note that in this context tameness is directly linked with our understanding of global symmetries in a theory that admits a UV completion with gravity. 
In fact, one of the best understood quantum gravity conjectures suggests that 
global symmetries must be either broken or gauged \cite{Banks:1988yz}. Applied to our situation, this means that either the full UV partition function 
does not have such a discrete symmetry group $G$, or that we should consider the theory on the quotient $\mathfrak{P}/G$. The above considerations treat exactly the gauged cases.

Let us now turn to an example with a broken global symmetry in which the tameness property is absent. Consider a theory with a quasi-periodic $\theta$-angle. For example take a theory 
with a $\theta \in \bbR$ appearing in the partition function as
$Z(\theta) = f(A \cos \theta + B \cos \alpha\theta)$ with $\alpha$ irrational. Such a dependence can arise, e.g., by considering a model with an effective scalar potential 
\begin{equation} \label{Veff-irrational}
   V_{\rm eff}= \tilde A \cos \theta + \tilde B \cos \alpha \theta\ .    
\end{equation}
The term $\cos \alpha \theta$ hereby breaks the periodicity $\theta \rightarrow \theta + 2\pi$. 
It was pointed out, for example, in \cite{BARDEEN1983445,Banks:1991mb,Blanco-Pillado:2004aap}, 
that such models have several interesting consequences. However, we note that such functions are never definable in any o-minimal structure due to the periodicity of the individual cosine-terms. This also applies to the linear plus cosine potential of
\cite{abbott1985mechanism}.
Despite their simplicity, we do not expect these theories to arise in a theory that admits a UV completion 
with gravity. To our knowledge, no realization of such a model has been found in string theory. While scalar potentials 
of the type \eqref{Veff-irrational} naturally arise in string theory, 
the coefficient $\alpha$ is always a rational number. 
It is interesting to point out that these constraints on the scalar potentials are reminiscent of the conditions arising from yet another quantum gravity conjecture, the distance conjecture, as recently discussed in \cite{Grimm:2022sbl}. 

So far we have been discussing tameness of observables as a function of continuous parameters.
Spaces of QFTs and of vacua also have discrete parameters, for example an integer valued WZW or Chern-Simons coupling,
or a quantized flux.  In part II we will discuss how this can be consistent with tameness.

\subsection{Action of the renormalization group} \label{RG-flow}

In this section we set out the hypothesis that renormalization group flow is tame.  
We will not make a precise conjecture, in part because the conditions we would need to impose are not clear and 
in part because defining the RG as precisely as we would need to do goes well beyond our scope here.
The following is meant to make the point that this question is very central and could be
studied in a precise way.

Recall that the RG is a flow on the space of cutoff QFTs, usually formulated as a system
of ordinary differential equations for the couplings $g^i$ of operators $O_i$,
\begin{equation} \label{eq:rgdef}
-\Lambda\frac{d}{d\Lambda} g^i = \beta^i(g) ,
\end{equation}
defined so that a joint variation of $\Lambda$ and $g^i$ preserves physical observables measured at energies below $\Lambda$.
The linearized RG is obtained by
evaluating $\partial \beta^i/\partial g^j$, which in an appropriate diagonalized basis yields the expansion 
\beq \label{beta-pert}
  \beta^i_{\rm pert}(g) = (d-\Delta_i) g^i + \cO(g^2)\ ,
\eeq
where $d$ is the space-time dimension $d$ and $\Delta_i$ are the operator dimensions. 
The higher order terms can be computed using perturbation theory. In general there can also be nonperturbative terms, but we have little to say about them at present.

While there is a great deal of physics here, let us simply regard Eq. \eqref{eq:rgdef} as a mathematical
definition and observe that it has two ingredients: 
a space of theories $\cT$ parameterized by $g^i$, and a vector field $\beta$
on this space.  While the process of renormalization involves many choices, it is geometric; different
choices of conventions, contact terms, {\it etc.} are related by diffeomorphisms on the space of
couplings.\footnote{ This is arguably a tautology as if we were to find choices which were not related by
diffeomorphisms, we could introduce additional geometric structures to make the framework covariant.
For example, the dilaton in the 2d sigma model can be motivated this way \cite{tseytlin_conformal_1986}.
}
Thus, the question ``is the RG tame''
becomes, are the different renormalization schemes of physical interest related by
tame diffeomorphisms, and is there a renormalization scheme in which the components of $\beta$
are tame functions?  If so, are the solutions $g^i(\Lambda)$ of the RG flow \eqref{eq:rgdef} tame?

The simplest situation to consider is the linear approximation in \eqref{beta-pert} with a beta function $(d-\Delta_i) g^i$. In this case $\beta(g)$ is trivially definable in $\bbR_{\rm alg}$ and we can ask if the solutions $g^i(\Lambda)$ are 
tame as well. A version of this question was studied in \cite{MillerVector}. What one finds is that $g^i(\Lambda)$ is only definable in an o-minimal structure, namely $\bbR_{\rm exp}$ if all $\Delta_i$ are real. In case some of the $\Delta_i$ are complex one necessarily leaves the o-minimal setting and can, for example, encounter spiraling solutions. In the RG context they are known as RG limit cycles, and a rather exotic phenomenon whose interpretation is still under debate, see e.g.~\cite{curtright_rg_2012,bulycheva_rg_2014} or, more recently, \cite{Jepsen:2020czw}. It remains open if one should allow for such situations in a well-defined class of QFTs. 

To study the more general situation with a non-trivial $\beta^i_{\rm pert}(g)$, we need to make sure that our statements are well-defined and hence specify the class of QFTs we are considering.
For now, let us take these to be asymptotically free theories with renormalizable Lagrangians depending on finitely
many fields.  This includes Yang-Mills theory in $d\le 4$, linear sigma models in $d=2$, and many other interesting
classes of theories.  It does not include effective field theories in the more general sense (so, with
nonrenormalizable couplings suppressed by appropriate powers of the cutoff).  In this class of QFTs, we
can define Eq.~\eqref{eq:rgdef} perturbatively, using the diagrammatic formalism of our earlier discussion.
Furthermore we have good reasons to think that the resulting series expansions are related to exact results,
at least for large $\Lambda$ and as asymptotic expansions.

Based on the results of \S \ref{sec:tameness_theorem}
it seems very plausible that the resulting $(\cT,\beta)$ would be $\bbR_{\rm an,exp}$-definable 
at every order in perturbation theory.  
Indeed one might at first think that it is tautologically so, because the $n$-loop contribution to $\beta$ 
is a polynomial in the couplings.  However this is not the case in the standard renormalization schemes as
each term is an {\it a priori} general function of the masses, which we are asserting is definable.
Furthermore there are an infinite set of equations analogous to Eq.~\eqref{eq:rgdef} which govern the
anomalous dimensions of higher dimension operators; these are related to the expansions of amplitudes Eq.~\eqref{defIlj}
in powers of external momenta and are definable as well.

We still face the problem of summing this expansion and somehow adding in any nonperturbative corrections,
but again the claim that the exact result is tame in some o-minimal structure 
(perhaps different from $\bbR_{\rm an,exp}$) looks sensible.
Suppose it were, what would it imply?

We expect that tameness will put constraints on the possible singular behaviors of QFT.
The argument is that -- from the RG point of view -- there are two ways one can get singularities:
either from actual singularities in $\beta$, or from taking the IR ($\Lambda\rightarrow 0$) limit of
the flow. Tameness of $\beta$ will constrain both possibilities.
In particular, if the o-minimal structure admits an analytic cell decomposition as discussed in \S \ref{definability} at least along the cells, which support a real analytic $\beta$, one can use the results on tame dynamical systems, e.g.~presented in  \cite{rolin_quasianalytic_2006}.
This should allow one to get mathematical constraints such as
tameness of observables.
One could then
try to understand the IR limit in terms of a drastically reduced number of fields, perhaps using
``integrating out'' which we discuss next.  

\subsubsection*{Integrating fields out}
\label{integrating-out}

In actual use of the RG, a second step is often taken.  Define a ``heavy'' field $\phi$ as one with mass $m$ 
greater than the cutoff, $m \gtrsim\Lambda$.  Then one can integrate out $\phi$, classically by solving for its
equation of motion and removing it from the action,
and quantum mechanically by going on to add the effects of the loops which involve it.
This produces a different $(\cT,\beta)$ depending only on the other ``light'' fields, but again satisfying the
defining property that the physical observables are the same as for the original theory.  One sometimes considers the inverse
operation of ``integrating fields in'' as well.

Is this step tame?  To properly ask this question we need to discuss the expectation values of scalar fields as well.
Let us denote the space parameterized by the scalar fields as $\cS$, so now we assume that $(\cT,\cS,\beta)$
are tame and consider integrating out $\phi$.
On the classical level, this amounts to restricting to the submanifold of $\cS$ defined by $\partial_\phi V(\phi)=0$,
followed by linear projection on $\cS$ (dropping the $\phi$ coordinate) and on $\cT$ (taking operators
which depend on $\phi$ and substituting its expectation value).  We would need to take quantum effects into
account as well.  Note that the definitions of $\cS$ and $\cT$ are coupled and a careful definition must
deal with this.

The classical integrating out step uses tameness in a rather direct way. To see this, we start by asserting that $V(\phi)$ is a definable function of the scalars $\phi^K$ spanning some definable field space $\mathcal{S}$. Assuming that $V(\phi)$ is sufficiently often differentiable,\footnote{Differentiability follows from 
definability, when excluding finitely many `smaller' subsets of $\mathcal{S}$ 
as made precise by using the $C^p$-cell decomposition discussed in \S\ref{definability}.} definability ensures that $V(\phi)$ has only finitely many maxima and minima.
We thus have finitely many solutions of the vacuum condition $\partial V/\partial \phi^\kappa_{\rm heavy}=0$, leading
to a finite set of resulting effective field theories.
Furthermore, since definability is preserved when taking derivatives, the condition 
defines a definable set in $\mathcal{S}_{\rm light} \subset \mathcal{S}$ spanned by the light fields. Since two definable sets intersect in a definable set, we conclude that all coupling functions remain definable when restricted to  $\mathcal{S}_{\rm light}$.

\subsubsection*{Exact renormalization group}

Can we extend this discussion to a general effective field theory, without assuming that it has a renormalizable
UV limit?  This would clearly be very important for applying these ideas in quantum gravity and string theory.
Furthermore a picture based on renormalizable UV limits suggests that the space of QFTs might have many
disconnected components, corresponding to the many such limits.  This goes contrary to the usual intuition
that components of theory space are connected unless there is some topological obstruction to it, such
as anomaly matching.  A definition which does not prefer renormalizable QFTs might avoid this problem.

There is an RG framework which can deal with general EFTs, the exact renormalization group.
Some representative works are
\cite{polchinski_renormalization_1984,morris_exact_1994,arnone_manifestly_2006,rosten_fundamentals_2010,morris_properties_2022,cotler_renormalization_2022} and \cite{costello2022renormalization} which makes a rigorous definition for perturbative gauge theory.
Without going into details, in this framework theory space $\cT$
is parameterized by the full action $S = \int \sum_i g^i O_i$ considered as a functional on field space.
One can then define Eq.~\eqref{eq:rgdef} as a functional differential equation.  Its linearization is
analogous to a heat equation and it is believed to have similar mathematical properties to this equation.
Since the heat equation and related nonlinear PDEs are mathematically relatively tractable,
this is a promising observation.
Indeed, the best understood example is the beta function for the $d=2$ nonlinear sigma model \cite{friedan1980nonlinear},
which is essentially Ricci flow (and indeed was the original inspiration for the mathematical study
of Ricci flow).

Another potential advantage of the exact RG is that, in return for the difficulties of working with an infinite dimensional
space of theories, the equation corresponding to Eq.~\eqref{eq:rgdef} could be simplified.  Indeed, there are arguments 
({\it e.g.}~\cite{rosten_fundamentals_2010} \S IV) that if
Eq.~\eqref{eq:rgdef} (for scalar field theories)
is rewritten as an equation for the exponentiated action $e^{-S}$, it becomes exactly
a linear heat equation!  This would certainly be a powerful statement if one could work with it.
We should also mention that another analog of Eq.~\eqref{eq:rgdef} can be derived
in the holographic RG \cite{de_boer_holographic_2000}.
Also interesting are recent connections to information theory
 \cite{cotler_renormalization_2022,stout_infinite_2022}.

So far as we know, the study of tameness 
of functional differential equations
is unexplored mathematical territory. 
One might ask, for example, if the set of all solutions occurring in these settings can be used to define a o-minimal structure, as introduced in \S \ref{definability}. Even simpler related questions such as 
``does the space of Ricci flat metrics define an o-minimal structure'' do not
seem to have precise formulations in the literature. 
We hope this subject will receive more study.

\subsection{Tameness in effective field theories and conformal field theories} \label{tame-EFTs} 

It is interesting to ask if one can introduce a well-defined notion of a parameterized 
set of QFTs such that we can inquire about the tameness of this set. This would then allow us to ask if the observables computed in this set are tame functions of the parameters or even on spacetime. In the following, we want to comment on the two classes of theories in which we expect that 
such tameness results can be established. Namely, we briefly discuss the set of 
effective theories compatible with quantum gravity, and the set of conformal field theories.
A more complete study, outlining a strategy to establish general tameness results in these classes of theories, will be presented in the upcoming work \cite{inProgress}.

In our studies so far, we 
have seen that tameness depends on the properties of the UV theory. In particular, we have seen 
in  \S \ref{nonpert-challenges} that it is easy to state UV Lagrangians that are neither tame functions of the fields nor of the parameters. For example, we saw that tameness is immediately violated in the presence of discrete global symmetries of infinite order. The latter are believed to be gauged or broken in quantum gravity. This can be viewed as an indication that tameness in effective theories is required in order to couple the theory to quantum gravity. 
This fact can be viewed as further evidence for a `Tameness conjecture' \cite{Grimm:2021vpn}, which was proposed in the spirit of the swampland program~\cite{Palti:2019pca,vanBeest:2021lhn} and gives a significant extension of previous finiteness conjectures \cite{douglas_statistics_2003,vafa_string_2005,acharya_finite_2006,douglas_spaces_2010,heckman_fine_2019,Hamada:2021yxy}.\footnote{This conjecture was originally based 
on the observation that all effective theories derived from string theory have strong tameness properties.}

To recall this conjecture, let us consider the set of Lagrangian effective theories with Einstein gravity that are valid at least up to some fixed cut-off energy scale $\Lambda$ and admit a completion with quantum gravity. The basic claim is that the Lagrangians of all such theories can be specified by sets and functions that are definable in some o-minimal structure $\bbR_{\text{EFT}d}$. 
Note that this statement requires to introducing an abstract notion of parameter space that collects all non-dynamical information about the effective theories. This can include constants appearing in the Lagrangian, e.g.~setting the strength of a coupling as in \S \ref{sec:tameness_theorem}, or even the number of fields that are considered. The conjecture also proposes the tameness of field spaces and all functions in the Lagrangian varying over it. 

Note that the o-minimal structure $\bbR_{\text{EFT}d}$ in these claims is not specified. 
In accordance with what we found for perturbative amplitudes in \S \ref{sec:tameness_theorem}
we might speculate that $\bbR_{\text{EFT}d}= \bbR_{\rm an,exp}$. However, we have noted in \S \ref{nonpert-challenges} that is might well be necessary to consider other structures. It is also reasonable to imagine that the choice of structure depends on $\Lambda$ and the amount of additional symmetries, such as supersymmetry, we demand on the theory. 

An immediate consistency check for the Tameness conjecture is provided by reevaluating tameness after lowering the cutoff $\Lambda$ in a tame effective theory.  
In this case, we need to study the renormalization group flow discussed in \S\ref{RG-flow} and potentially integrate out fields. We have seen in \S\ref{RG-flow} that classically tameness is indeed preserved. At the quantum level we have used the definability of loop amplitudes to show that at least in a renormalizable theory these steps are plausibly preserving tameness. In a nonrenormalizable effective theory, the precise renormalization procedure 
becomes relevant and a full treatment is beyond the scope of this work. However, it is tempting to speculate that actually the UV theory itself is tame even when it includes quantized gravity and that any sufficiently carefully extracted effective theory preserves this tameness.

Turning to the space of CFTs, it is clear that one must put some upper bound on the number of degrees of freedom
of the CFTs being considered, to have any hope for this to be tame.  The obvious quantities to bound are those
which decrease under RG flow, the central charges $c$ in $d=2$ and $a$ in $d=4$, and conjecturally the free energy
on $S^3$ in $d=3$.  The conjecture that sets of CFTs with such a bound are tame generalizes many conjectures in the literature,
such as the finiteness of Calabi-Yau $n$-folds.  But as we will see in part II, it is not true without
placing further conditions.  This is not inconsistent with the previous EFT conjecture as long as we accept (as is
generally believed) that not all CFTs are dual to theories of gravity on AdS.  The EFT conjecture furthermore suggests
that the subset of CFTs with AdS duals is tame, a conjecture we will examine as well.

\section{Conclusions} 

While physicists have learned to accept the many wild phenomena of quantum theories, the hope remains that at least 
the mathematical structure of these theories is tame and inherently geometric in nature. In this work we have 
shown that one can indeed formulate 
a general tameness principle, using the tame geometry built from o-minimal structures, that is common to many well-defined 
quantum theories. Concretely this means that physical observables are drawn from the special  
set of functions that are definable in an o-minimal structure. Such tame functions have strong finiteness 
properties, but nevertheless are sufficiently general to cover very complicated physical situations 
with singularities, runaway behaviour, or exponential dependence. This remarkable balance 
has a deep counterpart in mathematical logic in which mathematicians were aiming 
to find larger and larger o-minimal structures while preserving their central tameness property. 

A main result of our study was to establish that all $n$-loop amplitudes of a quantum field theory,
with finitely many fields and interactions, are definable in the o-minimal structure $\bbR_{\rm an,exp}$. 
This result followed from the fact that each such amplitude can be obtained by adding a finite number 
of Feynman integrals that can be related to period integrals of auxiliary geometries. The definability 
of period integrals in $\bbR_{\rm an,exp}$ then implies the statement. We note that definability 
holds for the real $n$-loop amplitudes as the functions of external momenta, coupling constants, 
and masses. The detour 
via a complex variety representation is thus not needed to state or use the final definability result. 
It can be also formulated as the observation that starting with a sufficiently tame quantum field theory Lagrangian, i.e.~a Lagrangian 
allowing for a perturbative treatment with finitely many fields and interactions, stays tame at 
the $n$-loop level. It does, however, not imply that tameness, namely the definability of the full 
amplitude in $\bbR_{\rm an,exp}$, is necessarily preserved when formally summing up all loops.  
While finite products and sums of definable functions are definable, this argument does clearly not 
extend to infinite sums. 

Establishing tameness results beyond perturbation theory is a very challenging task. 
In addition to addressing the full perturbative expansion 
also non-perturbative effects need to be included. 
Nevertheless we were able to show the full non-perturbative tameness of the partition 
function in several simple quantum theories in various dimensions for which it has 
been determined completely. While the definability of period integrals 
was again one of our main tools, it became apparent that one needs 
more general mathematical results when talking about a simple $\phi^4$ theory in zero dimensions or general quantum mechanical systems. This might also force us to introduce novel o-minimal structures replacing $\bbR_{\rm an,exp}$. In fact, we will suggest in the follow-up paper \cite{inProgress} that one should construct the structure associate to well-defined sets of QFTs and show that it is o-minimal.

We have argued that tameness will generally depend on the UV behaviour of the theory. In particular, we have seen 
that non-perturbative effects can naturally lead to periodic corrections that would violate tameness. In fact, any 
discrete symmetry of infinite order needs to be absent in the UV theory, which matches nicely with the expectation 
that all global symmetries in quantum gravity are gauged or broken. However, there are many other ways that tameness can 
be violated and there is no a priori argument against non-tame UV Lagrangians. In contrast, we expect that starting 
with a tame theory that the RG flow preserves tameness. We have gathered evidence for this idea by looking at the linear beta functions and leading perturbative corrections, which we know to be definable in $\bbR_{\rm an,exp}$. 
The resulting first order differential equations have been studied 
in the mathematics literature and we have highlighted this as an promising direction for further research. 
We have also argued that the integrating out process preserves tameness.

Tameness appears to arise in all known field theories that are obtained as 
a low energy effective action of string theory. This example-based evidence has led to
the Tameness conjecture for effective theories~\cite{Grimm:2021vpn} which asserts that 
effective theories valid below a fixed cut-off scale that can be consistently 
coupled to quantum gravity need to be tame. Our general arguments about the tameness of loop corrections indicate the self-consistency of this conjecture under lowering the cut-off scale. 
It would be desirable to go further in this direction by studying the full renormalization group flow.

In search for a fundamental principle which requires tameness, one tempting suggestion is to link tameness with logical decidability. Famously, G\"odel's theorems imply that there are undecidable statements in any axiom system formulated using the natural numbers and with arithmetics. These undecidablility statements are no longer true if one works with the real numbers \cite{Tarski}, and indeed tame geometry has much better decidability 
properties \cite{Marker1996ModelTA}.\footnote{This can be made much more precise when fixing the o-minimal structure under consideration. Firstly, all o-minimal structures 
are model complete \cite{VdDbook}. However, whether or not an o-minimal structure only yields decidable statements is a stronger condition. It was shown to be true for $\mathbb{R}_{\rm alg}$ \cite{Tarski}, and holds for $\mathbb{R}_{\rm exp}$ when assuming Schanuel's conjecture \cite{MacW}. 
} It is therefore appealing to conjecture that all statements about physical observables in tame quantum field theories are decidable.  Such an assertion would 
resolve some of the puzzles raised in Euclidean quantum gravity \cite{Geroch-Hartle,cmp/1104253849},  
condensed matter and statistical physics \cite{Cubitt2015UndecidabilityOT,Shiraishi2021UndecidabilityIQ}, and special quantum field theories \cite{Tachikawa:2022vsh}. 
We plan to expand on these observations in future work. 
It is interesting to inquire about the status of decidability within string theory and tame geometry 
might provide the best mathematical language to address these questions.\footnote{For a recent discussion of decidability in certain string compactifications, see e.g.~\cite{Cvetic:2010ky,Halverson:2019vmd}.}

In this work we have focused on establishing tameness properties of certain physical observables and did not touch much on the interesting implications it can have. To begin with, let us note that tame geometry allows to establish far reaching theorems previously only known within algebraic geometry. For example, we have recalled that tame sets and functions can be decomposed into finitely many cells. This fact can be used to associate topological invariants to sets and functions group them into equivalence classes. Furthermore, in many situations these can be represented by simplices \cite{VdDbook}. These ideas are readily applicable to physical settings and give new ways to distinguish theories on a fundamental level. More recently, remarkable mathematical advances show how powerful tame geometry is when combined with other structural criteria such 
as analyticity, see e.g.~\cite{MokPilaTsimerman,BT,GaoKlingler,Chiu,BaldiKlinglerUllmo,BTnew}.  
For example, the definability of periods together with their analyticity properties can be used 
to relate algebraic relations among them to special geometric symmetries of the setting. How this can be used to gain a deeper understanding of the relations and symmetries of Feynman amplitudes 
will be the topic of future work.

There are many directions in which this work can be extended.
One interesting direction would be to combine tameness with the use of resurgence in quantum field theory.  
Another is to study the tameness of spaces of conformal and quantum field theories and their observables,
which will be the subject of part II.

\subsubsection*{Acknowledgements}

We would like to thank Ofer Aharony, Ben Bakker, Matthias Gaberdiel, Matt Kerr, Bruno Klingler, Maxim Kontsevich, Eran Palti, Julio Parra-Martinez, Erik Plauschinn, 
Zohar Komargodski, Stefan Vandoren, and Mick van Vliet for useful discussions and comments. The research of TG and LS is supported, in part, by the Dutch Research Council (NWO) via a
Start-Up grant and a Vici grant.

\appendix

\section{Tameness of hypergeometric functions} \label{tame-hypergeometry}
In this appendix we review properties of the generalized hypergeometric functions emphasizing their tameness properties. The hypergeometric functions form a nice set of functions to point out the subtleties that arise when trying to establish definability results, as the functions contain both, examples and counterexamples, of tame functions. The generalized hypergeometric function is defined as the power series
\begin{equation}
    \label{hypdef}
    f(x)=\pFq{p}{F}{q}{\vec{a}}{\vec{b}}{x}=\frac{\prod\limits_{i=1}^{q}\Gamma[b_i]}{\prod\limits_{i=1}^{p}\Gamma[a_i]}\sum\limits_{n=0}^\infty \frac{\prod\limits_{i=1}^{p}\Gamma[a_i+n]}{\prod\limits_{i=1}^{q}\Gamma[b_i+n]}\frac{x^n}{\Gamma[n+1]}\;.
\end{equation}
The parameter vectors $\vec{a}$ and $\vec{b}$ have $p$ and $q$ entries respectively. In this section we will assume all parameters to be positive, as otherwise the functions are trivially tame.\footnote{For any integer $a_i<0$ the functions are polynomial and for any integer $b_i<0$ they vanish identically.} Many special functions, which are commonly appear in physical systems, are special cases of these hypergeometric functions. Here we only give a small group of examples:
\begin{align}
    \pFq{0}{F}{0}{}{}{x}&=e^{x}\;, \nonumber\\
    \pFq{1}{F}{0}{a}{}{x}&=\frac{1}{(1-x)^a}\;,\\
    \pFq{0}{F}{1}{}{1/2}{\frac{x^2}{4}}&=\cos(x)\;. \nonumber
\end{align}
where the entries are left empty when irrelevant for the expression. While the first two examples are obviously definable in $\mathbb{R}_{\rm{an,exp}}$ for $x\in \mathbb{R}$, the cosine function is only definable on a finite interval.
The convergence of the series \eqref{hypdef} depends on the relation between $p$ and $q$:
\begin{itemize}
    \item If $p<q+1$ the series converges for any value of $x$. The functions therefore reduce to restricted analytic functions when considered on any finite-length interval.\footnote{Note that one can set the function to, e.g., zero outside this interval, if one wants to define a function on all of $\bbR$.} Examples of this type are the sine and cosine functions. It is important to note that the restriction to a finite interval 
    excludes the essential singularity at infinity.
    \item The case $p=q=0$ for which $_0F_0(x)=e^{x}$ deserves a special emphasis. The function is of the type $p<q+1$ and therefore yields a restricted analytic function when considered on 
    a finite-length interval. Clearly, $e^x$ has an essential singularity at infinity. As discussed in \S \ref{definability}, it is a remarkable fact that one can construct o-minimal structures in which the exponential function on $\bbR$ is definable.
    \item If $p=q+1$, the series converges only for $|x|<1$. For $|x|>1$ the functions have to be analytically continued. The definability then depends on the exact properties of the monodromy groups around the boundaries. The periods of Calabi-Yau manifolds fall into this category, 
    e.g.~the period of an elliptic curve can be expressed in terms of
\begin{equation}
    \pFq{2}{F}{1}{1/2,1/2}{1}{x}=\frac{\pi}{2}K(x)\;,
\end{equation}
where $K(x)$ denotes the elliptic integral. As these are periods, the functions are definable. But in this case there is a more direct way to show the definability by a detailed study of the analytic continuation or the monodromy group. The definability of $K(x)$ and its derivative $\partial_x K(x)$ was proved in \cite{bianconi2016some} using this method.
    \item If $p>q+1$ the series diverges for any value of $x$ and  thus is an asymptotic series. In these cases the analytic continuation determines the whole function. For example, the simplest case of this class is given by $_2F_0(a,b,x)$, which can be expressed as
    \begin{align*}
    _2F_0(a,b,x)&=(-z)^{-a}U\Big(a,a-b-1,-\frac{1}{z}\Big) \\&
    = -\left(\frac{1}{z}\right)^{-a+b+2}\frac{\Gamma (a-b-2)} {\Gamma (a)} \pFq{1}{F}{1}{b+2}{b-a+3}{-\frac{1}{z}}\\&
    \ \ \ \, +\frac{\Gamma (-a+b+2) \,}{\Gamma (b)}\pFq{1}{F}{1}{a}{a-b-1}{-\frac{1}{z}}\;,
   \end{align*}
    where $U(a,b,x)$ is Tricomi's confluent hypergeometric function \cite{Tricomi1947-ew}. The analytic continuation is then given by functions of the type $p<q+1$ which where already discussed above.
\end{itemize}
While studying the definability case by case is feasible for low values of $p$ and $q$, this becomes rather involved for large values. But the generalized hypergeometric functions can be constructed recursively using a representation in terms of a generalized Euler integral, i.e.
\begin{equation}
\label{Eulerintegral}
    \pFq{p+1}{F}{q+1}{\vec{a},c}{\vec{b},d}{x}=\frac{\Gamma[d]}{\Gamma[c]\Gamma[d-c]}\int_0^1t^{c-1}(1-t)^{d-c-1}\pFq{p}{F}{q}{\vec{a}}{\vec{b}}{t x}\mathrm{d}t\;.
\end{equation}
Using this identity repeatedly allows the representation of the generalized $_pF_q$ function as a multiple integral over an algebraic function times either $_0F_0$, $_aF_0$ or $_0F_b$. If $p=q+1$, it reduces to
\begin{equation}
     \pFq{p+1}{F}{p}{\vec{a}}{\vec{b}}{x}=c\int_0^1\mathrm{d}t_1\int_0^1\mathrm{d}t_2\ldots \int_0^1\mathrm{d}t_p\;\omega\;\pFq{1}{F}{0}{a}{}{t_1t_2\ldots t_p x}\;,
\end{equation}
where the constant $c$ is a combination of gamma factors and $\omega$ is an algebraic function of the $t_i$. As
\begin{equation}
    \pFq{1}{F}{0}{a}{}{t_1t_2\ldots t_p x}=\frac{1}{(1-t_1t_2\ldots t_p x)^a}
\end{equation}
the integrand is an algebraic function. The hypercube domain is also definable, thus this kind of functions is definable, as expected from the period property. An important example of this type of functions are the polylogarithms. E.g.~the dilogarithm is given by
\begin{equation}
    {\rm Li}_2(x)=x\; \pFq{3}{F}{2}{1,1,1}{2,2}{x}=x \int_0^1 {\rm d} t_1\int_0^1{\rm d} t_2\; \frac{1}{1- t_1 t_2 x}\;.
\end{equation}
This can be generalized to any polylogarithm ${\rm Li}_n(x)$. For $n>0$ 
\begin{equation}
    {\rm Li}_n(x)=x\; \pFq{n+1}{F}{n}{1,1,\ldots,1}{2,\ldots,2}{x}=x \int_0^1 {\rm d} t_1\ldots \int_0^1{\rm d} t_n\; \frac{1}{1- t_1 t_2\ldots t_n x}\;.
\end{equation}
For $n\le 0$ the polylogarithms reduce to rational functions. Thus they are tame for any $n$. The polylogarithms show up in the partition functions of the topological A-model and non-critical M-theory, see \S\ref{sec:partition}. The $p=q$ case behaves differently. In this case the recursion leads to an integral of the form
\begin{equation}
     \pFq{p}{F}{p}{\vec{a}}{\vec{b}}{x}=c\int_0^1\mathrm{d}t_1\int_0^1\mathrm{d}t_2\ldots \int_0^1\mathrm{d}t_p\;\omega\; \pFq{0}{F}{0}{}{}{t_1t_2\ldots t_p x}\;.
\end{equation}
As $_0F_0(x)=e^{x}$, this integral has the form of an exponential period. The tameness of such functions is much less understood. The class of functions certainly contains examples which are not definable in $\mathbb{R}_{\rm{an,exp}}$ \cite{VANDENDRIES200161}. For example, the error function
\begin{equation}
    {\rm{erf}}(x)=\frac{2x}{\sqrt{\pi}}\pFq{1}{F}{1}{1/2}{3/2}{-x^2}\;,
\end{equation}
is an exponential period of this type but is not definable in $\mathbb{R}_{\rm{an,exp}}$. It is interesting to note that there exists a larger o-minimal structure $\mathbb{R}_{\rm Pfaff}$ of Pfaffian functions in which these kind of functions are definable \cite{Wilkie1999ATO,rolin_2008}. 

\section{An example: The Bubble Graph} \label{bubbleGraph}

\begin{figure}[h!]
    \centering
    \begin{tikzpicture}
\draw (1,0)--(2.5,0);
\draw (-2.5,0)--(-1,0);
\draw(0,0) circle (29pt);
\node at (-1.5,0.5) {$\vec{p}$};
\node at (1.5,0.5) {$\vec{p}$};
\node at (0,1.5) {$\vec{l}$};
\node at (0,-1.5) {$\vec{p}-\vec{l}$};
\end{tikzpicture}
    \caption{1-loop bubble Feynman diagram }
    \label{fig:feynmandiagram}
\end{figure}
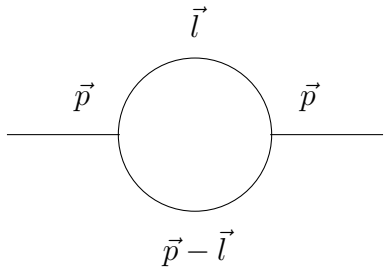

The construction relating Feynman integrals to periods is rather abstract. In this appendix we study a simple example of this construction, the 1-loop scalar bubble graph in $\phi^3$-theory, in detail. The Lagrangian of the theory is given by 
\begin{equation}
    L=-m^2 \phi^2+\lambda \phi^3\;.
\end{equation}
As this theory only has a single field and vertex, the mass gets only a single correction at the 1-loop level via the bubble or self-energy diagram shown in Figure  \ref{fig:feynmandiagram}.
The diagram depicted in Figure \ref{fig:feynmandiagram} corresponds to the Feynman integral
\begin{equation}
    I=-\frac{i}{2}\int\frac{\mathrm{d}^dl}{(2\pi)^4}\frac{1}{l^2+m^2-i\epsilon}\frac{1}{(p-l)^2+m^2-i\epsilon}\;.
\end{equation}
The integral consists out of two propagators, each appearing with exponent 1. Thus we have $v_1=v_2=1$ and $v=v_1+v_2=2$.
The first Symanzik polynomial is given by
\begin{equation}
    F=x_1^2+x_2^2+u x_1x_2\;.
\end{equation}
This follows from graph theoretic considerations, the details for the example are given in appendix \ref{appendix1}.
The resulting period integral corresponding to the maximal cut integral is given by
\begin{equation}
    I_{\rm cut}=\frac{1}{2\pi i}\int_{S_1}\frac{x_2\mathrm{d}x_1-x_1\mathrm{d}x_2}{\sqrt{z}(x_1^2+x_2^2)+x_1x_2}=\frac{2}{\sqrt{1-4z}} \;,
\end{equation}
where we have introduced the coordinate $z=1/u^2$.
The actual Feynman integral with the open contour is given by
\begin{equation}
\label{targetintegral1}
    I=\int_{x_2\ge 0}\frac{-\mathrm{d}x_2}{\sqrt{z}(1+x_2^2)+x_2}=\frac{2\mathrm{ArcTan}(\sqrt{4z-1})}{\sqrt{4z-1}}\;.
\end{equation}
In more general cases these period integrals cannot be directly evaluated and one needs another method of computing them. Following our idea of the main text, we interpret the polynomial in the denominator as the defining polynomial of a hypersurface with complex structure $u$. There exist several ways how to construct the GKZ system for this geometry, e.g.~by constructing the corresponding toric variety. The example has been worked out in \cite{Klemm:2019dbm}. Here we simply note that the $l$-vector relevant to this geometry is $l=(1,1,-2)$. Once one knows the $l$-vector, the holomorphic solution can be explicitly given \cite{Hosono:1993qy,Hosono:1994ax,Alvarez-Garcia:2020pxd,Alvarez-Garcia:2021mzv}. In the example the fundamental period is
\begin{equation}
\label{eq:1}
    \omega_0=\pFq{1}{F}{0}{\;\frac{1}{2}}{}{4z}=\frac{1}{\sqrt{1-4z}}
\end{equation}
As the GKZ system of this simple example is of order one this gives a complete basis of the periods. They are annihilated by the PF operator
\begin{equation}
\label{pfoperator}
    D=(1-4z)\theta-2z\;,
\end{equation}
where $\theta=z\partial_z$. The next step is to find the relativ periods, i.e.~one has to solve an inhomogeneous extension of the GKZ system.  To find the inhomogenity one acts with the operator \eqref{pfoperator} on the original Feynman integral, i.e.
\begin{equation}
    D\int_{x_2\ge 0}\frac{x_1\mathrm{d}x_2}{\sqrt{z}(1+x_2^2)+x_2}=-1\;.
\end{equation}
Thus one has to find a special solution for $  Df(z)=-1$. Luckily, this equation is solved by 
\begin{equation}
\label{inhomogeneuos}
    f(z)=-\frac{2 \arctanh{\left(\sqrt{1-4 z}\right)}}{\sqrt{1-4 z}}.
\end{equation}
This is equivalent to the solution to the Feynman integral \eqref{targetintegral1}. While this method works, the process of working out the inhomogeneties becomes rather involved in examples with more moduli. And even more importantly, we do not know if the solutions to the inhomogeneuos equations are definable. To remedy this situation, we note that this relative period is the same as the second period of the $\epsilon$ deformed GKZ system. The hypergeometric function in \eqref{eq:1} can be written as a power series
\begin{equation}
    \pFq{1}{F}{0}{\;\frac{1}{2}}{}{4z}=\sum\limits_{n=0}^{\infty} \frac{(2n)!}{(n!)^2}z^n\;.
\end{equation}
Using the usual Frobenius trick by replacing $n\rightarrow n+\epsilon$ to obtain the deformed system one gets the hypergeometric function
\begin{align}
\label{epsilonexpan}
   &z^{\epsilon }\frac{ (2\epsilon)! }{(\epsilon!)^2}\; _2F_1\left(1,\epsilon +\frac{1}{2};\epsilon +1;4 z\right)=\\&\frac{1}{\sqrt{1-4z}}-\frac{\log (4z)-2 \log \left(\sqrt{1-4 z}+1\right)}{\sqrt{1-4z}} \epsilon+\mathcal{O}(\epsilon^2),
\end{align}
which can be evaluated using HypExp2\cite{Huber:2007dx}. With some algebra one can see that the coefficient of $\epsilon$ in this expansion is equivalent to the inhomogeneous solution \eqref{inhomogeneuos}.

The coefficients in this expansion correspond to the periods of the underlying manifold and they agree exactly with the boundary computation. But they are only periods up to an order equal to the dimension of the manifold. Due to the relative period we exceed the dimension by 1, leading to a semi-period.  In general it is not known if semi-periods are definable. But the hypergeometric function in \eqref{epsilonexpan} is also the period of an elliptic curve. Thus the relative period of a point can be expressed as the period of an elliptic curve. In the language of Feynman integrals, this corresponds to the statement that the bubble diagram appears as a subgraph of the sunset graph.
The holomorphic solution of the sunset graph can be obtained in the same way as for the bubble diagram and is given as
\begin{align}
    \omega=\sum_{\ov{m}\ge 0}\frac{ \Gamma \left(m_1+m_2+m_3+m_4+1\right)}{\Gamma \left(m_1+1\right) \Gamma \left(-m_1+m_2+1\right) \Gamma \left(m_1-m_2+m_3+1\right)}\cdot \\
    \frac{1}{\Gamma \left(m_3-m_4+1\right) \Gamma \left(m_4+1\right) \Gamma
   \left(m_2-m_3+m_4+1\right)}
    z_1^{m_1}z_2^{m_2}z_3^{m_3}z_4^{m_4}\;,
\end{align}
where we abbreviated $\ov{m}=\{m_1,m_2,m_3,m_4\}$.
This period has 4 parameters, but the physical parameter space is spanned by only 3 parameters. The physical subspace is given by $z_1=z_4$.
Due to the symmetry of the sunset graph, the bubble graph arises in any limit where one of the three remaining $z_i$ vanishes. As an example we choose here $z_3=0$. This is a boundary limit, so the period will be a mixed period at this point. With this choice, the period simplifies to
\begin{equation}
  \omega=  \sum_{m_1\ge 0\;,m_2\ge 0}\frac{ \Gamma \left(m_1+m_2+1\right)}{\Gamma \left(m_1+1\right) \Gamma \left(m_1-m_2+1\right) \Gamma \left(m_2+1\right) \Gamma \left(-m_1+m_2+1\right)}z_1^{m_1} z_2^{m_2} \;.
\end{equation}
Due to the $\Gamma$ functions in the denominator, only terms with $m_1=m_2$ will contribute to the sum, simplifying this further to:
\begin{equation}
    \omega=\sum_{m_1\ge 0}\frac{\Gamma \left(2 m_1+1\right)}{\Gamma \left(m_1+1\right){}^2}z_1^{m_1} z_2^{m_1}=\frac{1}{\sqrt{1-4z_1z_2}}\;.
\end{equation}
Setting $z_1z_2=z$, this becomes exactly the fundamental period of the bubble graph. The same argument with the $\epsilon$ deformation holds as before, but now the second term in the expansion is still a period. Thus the relative period of the bubble graph arises as a mixed period of the sunset graph. The same structure appears at each loop level. The relative period of the sunset graph can be obtained by exceeding the order of the $\epsilon$ expansion by 1. This period can then be obtained by going to a boundary of the 3-loop banana graph.

\section{Graph Polynomials}
\label{appendix1}
In deriving the definability of Feynman integrals we used the Lee-Pomeransky representation. To make this paper more self-contained 
we will review the graph theory necessary for the definition of the Symanzik polynomials. We closely follow \cite{Bogner:2010kv}.

A Feynman diagram is a connected graph consisting out of a set of internal edges $E=\{e_1,e_2,\ldots,e_n\}$ representing the propagators as well as a set of vertices $\{v_1,v_2,\ldots,v_r\}$. The graph has $\ell=n-r+1$ loops. To define the polynomials one first introduces spanning trees. A spanning tree is a connected graph without loops which includes all vertices of the diagram. These can always be obtained from the original graph by removing $\ell$ edges. Furthermore, a k-forest is a graph without loops including all vertices consisting out of k connected components. A 1-forest is thus given by a spanning tree. In general, k-forests are obtained by removing $k+\ell-1$ edges from the original graph.

k-forests are not unique, for each graph and $k$ there exist several k-forests. The set of all k-forests is denoted $\mathcal{T}_k$ and its elements $(T_1,T_2,\ldots,T_k)$. The $T_i$ denote the k spanning trees, i.e.~the connected components, of the forest.

With these definitions we can define the graph polynomials as 
 \begin{align}
     U&=\sum_{(T_1)\in \mathcal{T}_1}\prod_{e_i\notin T_1}x_i\ ,\\
     F&=-\sum_{(T_1,T_2)\in \mathcal{T}_2}\left(\prod_{e_i\notin T_1,T_2}x_i\right)\left(\sum_{e_i\in T_1}\sum_{e_j\in T_2}p_i\cdot p_j\right)+U\sum_{e_i\in E}x_i m_i^2\ .
 \end{align}
 
 For the example of the bubble graph discussed in appendix \ref{bubbleGraph} we have $E=\{e_1,e_2\}$, $n=r=2$ and $l=1$. The 1-forests or spanning trees are obtained by removing any of the two edges, i.e.~they consist out of the two vertices and one edge. For the 2-forests 2 edges have to be removed, thus there is a single 2-forest consisting  out of the vertices and no edges. The polynomials then become
 \begin{align}
     U&=x_1+x_2\\
     F&=x_1x_2(p_1\cdot p_2)+(x_1+x_2)(x_1 m_1^2+x_2 m_2^2)
 \end{align}
 The product $p_1\cdot p_2=p^2$ can be expressed in terms of a single variable due to momentum conservation and by rescaling the integration variable. Note that the form of the polynomial G relevant for the Lee-Pomeransky representation is closely related to $F$,
\begin{equation}
    G=U+F=x_1x_2(p_1\cdot p_2)+(x_1+x_2)(x_1 m_1^2+x_2 m_2^2+1)\,
\end{equation}
the only difference being the additional $+1$ in the last term. To arrive at the expression in the main text the coordinates are rescaled as $x_i\rightarrow \frac{x_i}{m_i}$ resulting in 
 \begin{equation}
    G=x_1x_2\frac{(p^2+m_1^2+m_2^2)}{m_1m_2}+x_1^2+x_2^2+x_1+x_2\;.
\end{equation}
The expression for $F$ in the previous section then follows by defining $u=(p^2+m_1^2+m_2^2)/(m_1 m_2)$.
As $G$ is a sum of polynomials of different degree, it is a non-homogeneous polynomial. This is remedied by adding an additional coordinate, $x_3$ in this case, such that the polynom becomes homogeneous. The integral over the loop momentum has thus been reinterpreted as a period integral over a hypersurface in $\mathbb{P}^2$, where the projective freedom is used to fix the $x_3$ coordinate to 1. Note that this leaves two integrations to be performed.
 If one would instead use the representation via the  Symanzik parameterization, one would end up with a single integral due to the $\delta$ function in \eqref{integral2}. In terms of the polynomials one can see the same effect, as the polynomial F is already homogeneous. This allows the use of the rescaling freedom to eliminate an additional variable, effectively reducing the integrals to be performed by 1, rendering this representation more efficient for the computation of l-loop banana integrals. But for general statements about amplitudes we have to rely on the Lee-Pomeransky representation.

\bibliography{literature}
\bibliographystyle{utphys}
\end{document}